%% file: mainbody.tex
\pdfoutput=1
\documentclass[journal]{IEEEtran}

\usepackage{cite}
\usepackage{amsmath,amssymb,amsfonts}
\usepackage{algorithmic}
\usepackage{graphicx}
\usepackage{textcomp}
\usepackage{xcolor}

\usepackage{lineno,hyperref}
\usepackage{siunitx}
\usepackage{amsmath}
\usepackage{amsmath,bm}
\usepackage{xcolor}
\usepackage{multirow}
\usepackage{comment}
\usepackage{makecell}
\usepackage{graphicx}

\usepackage{array}
\usepackage{mdwmath}
\usepackage{mdwtab}
\usepackage{eqparbox}
\usepackage{url}

\modulolinenumbers[5]

\newcommand{\RomanNumeralCaps}[1]
    {\MakeUppercase{\romannumeral #1}}

\DeclareMathOperator*{\argmin}{arg\,min}

\begin{document}

\title{High-Performance Permanent Magnet Array Design by a Fast Genetic Algorithm (GA)-based Optimization for Low-Field Portable MRI}

\author{\IEEEauthorblockN{Ting-Ou LIANG$^a$, Yan Hao KOH$^a$, Tie QIU$^a$, Erping LI$^b$, Wenwei YU$^c$, Shao Ying HUANG$^{a,d}$ } \\
\IEEEauthorblockA{$a$ \textit{Singapore University of Technology and Design, 8 Somapah Road, Singapore 487372}} \\
\IEEEauthorblockA{$b$ \textit{Zhejiang University, Hangzhou, Zhejiang Province, China}} \\
\IEEEauthorblockA{$c$ \textit{Center for Frontier Medical Engineering, Chiba University,\\Inage Ku, Yayoi Cho, 1-33, Chiba, 263-8522, Japan}} \\
\IEEEauthorblockA{$d$ \textit{Department of Surgery, Yong Loo Lin School of Medicine, National University of Singapore,\\IE Kent Ridge Road, Singapore 119228}}
}

\maketitle

\begin{abstract}
A permanent magnet array (PMA) is a preferred source of magnetic field for body-part-dedicated low-field ($<$\,0.5\,T) portable magnetic resonance imaging (MRI) because it has a small footprint, no power consumption, and no need for a cooling system. The current popular PMA is limited by the field strength to be below 100\,mT and the transversal field direction where advanced technologies developed for the long-bore MRI systems (e.g., multi-channel techniques) cannot be applied. 
In this paper, a sparse high-performance PMA is proposed based on inward-outward ring pairs and using magnet blocks that can be bought off the shelf, targeting on portable head imaging. Through a fast genetic algorithm (GA)-based optimization, the proposed PMA has a longitudinal magnetic field with an average field strength of 111.40\,mT and a monotonic field pattern with inhomogeneity of 10.57\,mT (an RF bandwidth of $<$\,10\%) within a Field of View (FoV) of 20\,cm in diameter and 4.5\,cm in length. The field generated by the design was validated using analytic calculations and numerical simulations. The field can be used to supply gradients in one direction working with gradient coils in the other two directions, or can be rotated to encode signals for imaging with an axial selection.
The encoding capability of the designed PMA was examined through checking the quality of the simulated reconstructed images.
When it is used for encoding, the field pattern favors imaging with good quality, which even outperforms a linear pattern. The magnet array is 57.91\,cm wide, 38\,cm long, has a 5-gauss range of 87\,$\times$\,87\,$\times$\,104\,cm$^3$, allowing an operation in a small space. It weights 126.08\,kg, comprising of a stationary main array that supplies a homogeneous fields with high field strength, and a light rotatable sub-gradient-array (5.12\,kg) that supplies a monotonic field for signal encoding. It has a magnetic field generation efficiency of 0.88\,mT/kg which is the highest among sparse PMAs that offer a monotonic field pattern.
The force each magnet experiences in the design was calculated and the feasibility of a physical implementation is examined. The design can offer an increased field strength thus an increased signal-to-noise ratio (SNR). It has a longitudinal field direction that allows the applications of technologies developed for a long-bore system, such as surface coils and multi-channel technology without compromising coil efficiency. 
This proposed PMA can be a promising alternative to supply the main field and gradient fields combined for dedicated portable MRI.
\end{abstract}


\IEEEpeerreviewmaketitle


\section{Introduction}\label{sec:intr}

Magnetic resonance imaging (MRI) offers good soft tissue contrast and has no ionizing radiation. A traditional MRI system is expensive, bulky, and stationary. It is hard to provide ``point-of-care'' and timely diagnosis, e.g., those in an ambulance, a field medical tent, or an intensive care unit (ICU). In recent years, portable MRI is obtaining increasing attention in both the academia and the industry\,\cite{hyperfine}. To achieve portability, the body-part-dedicated concepts helps to significantly reduce the physical size of the system. Meanwhile, permanent magnet array (PMA) becomes a popular options to supply the magnetic fields for imaging. 
A PMA has no power consumption and is low-cost. Compared to a superconducting magnet, a PMA does not need sophisticated cooling systems although the field strength is relatively low; compared to an electromagnet, it can supply much higher field strength and has no heat dissipation.
A PMA can be applied to supply a homogeneous fields, or gradient fields, or a combination of the above in an MRI system.
To generate a homogeneous field for MRI, \textit{in situ} PMAs, which have imaging taken place inside the array, can be used\,\cite{huang2019_iMRI}. There are mainly two types of \textit{in situ} PMAs, one is the dipolar magnet array (e.g. C-shaped\,\cite{Cheng2001_CShape} and H-shaped PMA) which is a magnet circuit consisting of two poles (with aggregated magnets) and an iron yoke. In a dipolar magnet array, the space between the two poles is used for imaging\,\cite{siemens}. The other type is a cylindrical magnet array, for example, a Halbach array\,\cite{halbach1980design} with tranversal fields, or an inward-outward (IO) ring pair with longitudinal fields\,\cite{Nishino1983singleRing,Miyajima1985Ring_pair,aubert1994permanent,ren2018design} where imaging is done in the bore. In the literature, a comprehensive review on permanent magnet and PMAs for MRI is presented\,\cite{huang2019_iMRI}.
To achieve portability for a PMA-based MRI system, when the homogeneity of the magnetic field has to be maintained, the size of a PMA has to be reduced by either reducing the field of view (FoV) or lowering the field strength. For example, when a C-shaped PMA is scaled down to a table-top size, the FoV has to be shrunk to a volume of 1.27$\,\times\,$1.27$\,\times\,$1.90 cm$^3$\,\cite{esparza1998low} which is not practical to image {most of body parts}. For a Halbach array, if the FoV is 3\,cm diameter of cylindrical volume, an average field strength of 220\,mT with 11 part per million (ppm) can be obtained\,\cite{Danieli2009}. However, when the FoV is increased to 20\,cm diameter of spherical volume (DSV), as reported by   
a recent publication\,\cite{OReilly2019LargeHalbach}, the magnetic field obtained is average at 50.64\,\si{\milli\tesla} with a homogeneity of 2,400\,ppm.

When the requirement on homogeneity of magnetic field is relaxed, a relatively large FoV can be obtained with a relatively small and/or light magnet array. A less homogeneous magnetic field can be used as a combination of the main fields and gradient fields for MRI\,\cite{cooley2015} which is called a spatial encoding magnetic field (SEM), borrowing from PatLoc (\textbf{Pa}rallel imaging \textbf{t}echnique using \textbf{Loc}alized gradients) imaging\,\cite{hennig2008parallel}. This gradient field supplied by a PMA may not be linear, as a result, the number of electrical gradient coils can be reduced or even be reduced to zero. One note is that SEM in a PMA system is a combination of the main and gradient fields whereas SEM in PatLoc is only the nonlinear gradient fields.
PMAs used to supply SEMs for MR imaging\,\cite{cooley2015,ren2015magnet} have their unique characteristics, and signature impacts on different aspects of the system, e.g. the pulse sequence, the image reconstruction and analysis, the RF system including the control circuit and RF coils, which is documented in\,\cite{jia2019effects}.
In terms of physical implementations, a long sparse Halbach magnet array was designed to provide a dipolar SEM with a quadrupolar pattern, pointing on the traversal plane, rotated in the $\theta-$direction for head imaging\,\cite{cooley2015}.
In\,\cite{ren2015magnet}, a short Halbach array was proposed for 2D head imaging. For the SEM supplied by a Halbach array, within a FoV of a head size (i.e. a DSV of 20\,cm), the pattern has certain symmetry with respect to the center of the cylinder, which causes non-bijectiveness when it is used to encode signals. Moreover, it shows low and even zero gradient in the central region. For the former issue, surface loop coils were introduced as receive coils on the wall of the cylinder which has a decaying coil sensitivity (away from the coil) to encode the amplitude of the signal, so as to eliminate the ambiguity\,\cite{Kelton1989_multi_Rx_coil,cooley2015two}. However, the remedy induces the next issue. The $B_1$-field from the receive coils points on the transversal direction, which is the same as the SEM supplied by the magnet. When the magnet rotates, it is not perpendicular to the SEM periodically and the efficiency of the system is compromised\,\cite{jia2019effects}.

To address this problem, IO ring-pair PMA\,\cite{ren2018design,ren2019irregular} were proposed to offer an SEM in the longitudinal direction for head imaging. They were re-designed based on a single IO ring pair that were proposed by E. Nishino\,\cite{Nishino1983singleRing}/G. Miyajima\,\cite{Miyajima1985Ring_pair} back in the 80's, to have intense and homogeneous magnetic fields for NMR/MRI. 
In the 90's, G. Aubert proposed to superpose multiple IO ring pairs of different sizes, constructing a cylindrical PMA to obtain a homogeneous field within a 40\,cm bore\,\cite{aubert1991cylindrical}.
In\,\cite{ren2018design}, an IO ring-pair aggregate consists of IO rings with an inner radius tapered outside in was proposed. It supplies a concentric field pattern which has a monotonic gradient in the $\rho-$direction. The downside of this design is that the gradient in the $\phi-$direction is zero. In\,\cite{ren2019irregular}, an IO ring-pair aggregate is further discretized along the $\phi-$direction and it generates a monotonic field pattern. It is shown that this monotonic field, when being rotated to encode signals, can provide a higher spatial resolution to a reconstructed image with fewer artifacts and more recognized features compared to a quadrupolar\,\cite{cooley2015two} or a concentric pattern\,\cite{ren2018design}.
Although an irregular IO ring-pair aggregate\,\cite{ren2019irregular} shows good signal encoding capability for imaging, it has relatively high inhomogeneity which may not work with the RF system (spectrometer and RF coils), and it is heavy. Moreover, the design consists of ring-segments (fan-shaped blocks) with different inner and outer diameters. It is hard to implement by using permanent magnet blocks off the shelf\,\cite{zhang2017advances}.

In a MRI system using a cylindrical PMA, e.g., the Halbach array or the IO ring-pair array, to supply SEM, if the gradient is linear, the PMA can be rotated for projection imaging or working with another two graident coils in the other two orthogonal directions for Fourier imaging\,\cite{cooley2021Nature_BE}. If the gradient is nonlinear, it can only be rotated for MR imaging\,\cite{cooley2015,ren2015magnet,ren2018design,ren2019irregular}.
When a rotation is needed for imaging in an MRI system, the rotation error effects the image quality\,\cite{jia2019effects,gong2020}. 
If the whole array need to be rotated where the array is usually heavy ($>$\,45\,kg), a powerful motor is needed. Moreover, a heavier array could lead to a heavier load of the motor, and thus a higher the mechanical rotational error. Lowering the load of the rotation motor in such an MRI system helps to lower the mechanical rotational error, which greatly helps to improve the quality of a reconstructed image, besides leading to a low power consumption.

\input{fig_3D2D_view.tex}

For the optimization of a PMA, genetic algorithm (GA) has been a widely used tool. In\,\cite{OReilly2019LargeHalbach,ren2018design}, it was used to optimize the PMA to obtain a homogeneous field pattern in the FoV for a Halbach array and an IO ring-pair PMA, respectively. In\,\cite{cooley2018design,ren2019irregular}, it was used to as an optimization tool for a Halbach array and an IO ring-pair PMA to obtain a monotonic field patter, respectively.

In this paper, an implementable and high-performance sparse cylindrical inward-outward (IO) ring-pair PMA is proposed. It generates a strong magnetic field along the axial direction with a monotonic field pattern for signal encoding for head imaging. It consists of a main array that supplies a strong relatively homogeneous magnetic field and a light (5.12\,kg) and a rotatable sub-gradient array that supplies a monotonic field pattern for signal encoding. It has an average field strength of 111.40\,mT and an inhomogeneity of 10.57\,mT within a FoV of 20\,cm in diameter and 4.5\,cm in length, which corresponds to a RF bandwidth of $<$\,10\%. All parts of the proposed array consists of magnet blocks with grade N52, N50, N48, or N45 that can be purchased off the shelf, which eases a physical implementation. 
A fast Genetic algorithm (GA) was used for the optimization and a in-house-built accelerated magnet simulator, {``MagCraft''}, was used for a forward calculation, to calculate the magnetic fields of magnet blocks that make up of the array.
The design was validated using analytic calculations and numerical simulations. The force among the magnets in the proposed array was checked for the feasibility of a physical assembly. 
With an average field strength higher than 110\,mT and a monotonic field pattern for encoding, the proposed magnet array leads to an improved MR image quality compared to other existing PMAs, which is numerically demonstrated. It can offer a high SNR to an MRI system, allows the application of MRI technologies developed for a longitudinal main field. It can be a promising alternative to supply the main field and gradient fields for body-part-dedicated MRI.

The design methods and optimization of each part of the PMA are elaborated in Section\,\ref{sec:methods}. Field distributions in the targeted FoV of the proposed PMA as well as those of each sub-array are presented in Section\,\ref{sec:results}. In the same section, their imaging encoding capabilities were examined by checking the image quality when the field was rotated and applied as a combination of the main field and gradient field for imaging. A few other issues, the 5-Gauss region, the force-related issues (force experienced by the magnet blocks and sub-arrays, and that by the blocks during assembly), and temperature drift of magnetic fields, are presented and discussed in Section\,\ref{sec:discussions}. At the end, this paper is concluded in Section\,\ref{sec:conclusion}. 

\input{fig_magnet_block_current_model.tex}

\section{Methods}
\label{sec:methods}

A high-performance IO-ring-based magnet array is proposed. It is shown in Fig.\,\ref{fig:array}. It is a result of a physics-guided optimization, using a fast GA. 
A physics-guided GA optimization consists of two steps. Step\,1 is the physics-guided design, a design process that is guided by the physics on magnets and magnet array which generates constraints on the design parameters. These constraints are passed to a GA optimization at Step\,2 to confine the searching landscape, which highers the chance of obtaining an optimal solution and accelerates the process. 
In this design, two types of ring pairs are used as the bases, one is the IO-ring type of ring pair where the magnet blocks have their polarizations pointing outward or inward, as shown in Fig.\,\ref{fig:block_current_model}\,(a), and the other is the parallel type of ring pair where the polarizations are pointing along the axial direction, as shown in Fig.\,\ref{fig:block_current_model}\,(b). The fast GA is enabled by a in-house-built accelerated magnet simulator, {``MagCraft''}\,\cite{supplement}.

\input{fig_3D_overview.tex}

\subsection{Design Objectives}
\label{subsec:design objectives}

The design target is a cylindrical PMA that supplies a magnetic field in the axial direction (set to be the $z$-direction).
The advantage of having fields in the $z$-direction is that it allows the use of a variety of high-efficiency transverse RF coils\,\cite{mispelter2015book}, and an easy incorporation to other MRI techniques, such as multi-channel techniques and pre-polarization techniques\,\cite{lurie2010paper}.
The FoV is a cylinder with a diameter, $d_\text{FoV}$, of 20\,cm and a thickness of 4.5\,cm, as shown in Fig.\,\ref{fig:3D_overview}.
For a good image quality, it is aimed to have a magnet array to supply a magnetic field that possesses the following characteristics, a high average magnetic field strength so as to have a high SNR, a pattern with ``good encoding capability'' that leads to good quality images when it is rotated to encode signals, and an inhomogeneity that are kept in a range working with the bandwidth of the RF coils and other RF components of the system. 
A pattern with ``good encoding capability'' means a monotonic field pattern to have a sufficient coverage in the local $k$-space when the field rotates\,\cite{jia2019effects}, and a high gradient globally and locally in the FoV to have a high spatial resolution.
Specifically, the design objectives were set to be an average field strength of greater than 100\,mT, an inhomogeneity of not more than 10\% of the average field strength (that corresponds to an RF bandwidth of not more than 10\%), and a monotonic field pattern.

\subsection{Overview of the Design}
\label{subsec:design overview}

Fig.\,\ref{fig:array} shows the proposed magnet array design. It consists of two coaxial sub-arrays, the main array (a combination of the base array and the shimming array that are coaxial) that provides a strong and relatively homogeneous ${B}_0$-field ($\textbf{B}_0$), and the sub-gradient array that supplies the spatial encoding magnetic field (SEM, denoted as $\textbf{G}^\mu_\text{SEM}$), both in the $z$-direction.
Therefore, the total magnetic field generated by the proposed magnet array, $\textbf{B}^\mu_\text{SEM}(\textbf{r})$, can be expressed as,
\begin{equation}\label{eq:SEM}
\centering
\textbf{B}^\mu_\text{SEM}(\textbf{r}) = \textbf{B}_0+\textbf{G}^\mu_\text{SEM}(\textbf{r})
\end{equation}
where $\textbf{r}$ is the position vector in the FoV. The bold letter in this content represents a vector. 
The proposed magnet array is sparse, consisting of magnet blocks.
Each sub-array of the proposed magnet array consists of sparse IO ring pair(s) and/or sparse parallel ring pair(s), as shown in Fig.\,\ref{fig:block_current_model}\,(a) and Fig.\,\ref{fig:block_current_model}\,(b), respectively. 
The sparsity of the magnet ring pairs is to facilitate a physical implementation and to lower the weight. 
The main array is stationary and the sub-gradient array can be rotated for signal encoding if it is needed. Thus, the former can be heavy whereas the latter was designed to be light to ease a rotation.

The main array is a shimmed IO ring pair structure. As shown in Fig.\,\ref{fig:array}, it consists of a sparse IO ring pair (called base array, in grey, supplying $B_\text{base}$ in the z-direction) and a cylindrical shimming array (in red, supplying $B_\text{shim}$ in the z-direction), leading to $B_0$\,=\,$B_\text{base}$\,+\,$B_\text{shim}$.
The base array and the shimming array has an inner radius of $R'_\text{base}$ and $R'_\text{shim}$, respectively, where $2R'_\text{base}\,>\,2R'_\text{shim}\,>\,d_\text{FoV}$.
For the base array of the main array, it is an IO ring pair which consists of big magnetic blocks in the front and the rear rings having the polarization pointing radially outward and inward, respectively.
The shimming array is a combination of the IO-type ring pairs and the parallel-type ones that are arranged with a sequence along the $\pm z$-direction and consists of small magnet blocks.

The sub-gradient array (in blue and green in Fig.\,\ref{fig:array}) is cylindrical. As shown in Fig.\,\ref{fig:array}, it is located between the two rings of the base array along the $z$-direction, and outside the shimming array in the radial direction. As shown in Fig.\,\ref{fig:array}\,(c), it is an array consisting of two cylindrical arrays (i.e., the green one and the blue one with a radius of $R'_{g}$, and $R'_{g}\,>\,R'_\text{shim}$) with an offset of the axes along the $x$-direction by $2\Delta x$. For each array of the sub-gradient array, it is composed of parallel ring pairs arranged along the $\pm z$-direction.
The details of the designs of the sub-arrays are presented in Section\,\ref{subsec:design_op_grad}. 

\input{fig_flowchart.tex}

A physics-guided fast GA optimization was applied to the design.
Fig.\,\ref{fig:flowchart} shows the flow of the optimization.
As shown in Fig.\,\ref{fig:flowchart}, there are two optimization blocks, the red one for the main array and the blue one for the sub-gradient array. In each block, there are the physics-guided design process at Step\,1 and a following fast GA optimization at Step\,2. 
Step\,1 is guided by the physics of magnets, narrows the ranges of the design parameters, and generates design constraints to confine the searching landscape at Step\,2. Step\,1 consists of pre-design inspections and design synthesizers. At Step\,2, GA with accelerated forward magnetic field calculation was used to find the optimal dimensions, locations, orientation, and polarization of the magnet blocks to meet the design objectives. 
Between the two optimization blocks in Fig.\,\ref{fig:flowchart}, the optimized main field $B_0$ from the red block is applied to each step in the blue block. It is used to refine the constraints of the gradient field $G$ in Step\,1 of the sub-gradient array optimization, and meanwhile, it is input to Step\,2 for the forward calculation of the total field for the optimization of the sub-gradient array. After the optimizations for the main and the sub-gradient arrays, the field outputs from the red and the blue blocks are combined, ($B_0$\,+\,$G$), and applied to image a digital phantom to check the performance of the combined field when it is used as encoding fields for imaging\,\cite{gong2019local_k}. The image quality is checked by using structural similarity index method (SSIM). If it is satisfactory, all the outputs from the two blocks, $\Bar{\psi}^\text{op}_\text{main}$, $\Bar{\psi}^\text{op}_\text{subgrad}$, and ($B_0$\,+\,$G$) are outputs. Otherwise, the result is fed back to the design synthesizer of Step\,1 of both the red and blue blocks to re-trigger a new round of optimization by refining the constraints. 

For each GA optimization at Step\,2 of the red or blue block, a population with $N$ individuals is initialized. Each individual is an array of variables of the targeted sub-array, $\Bar{\psi}_\text{base/shim/subgrad}$ or $\Bar{\psi}_\text{main}$ with suggested ranges based on the pre-design inspections and design synthesizer at Step\,1. Based on the variables, a fast forward calculation of the magnetic field (outlined in green in the flow chart in Fig.\,\ref{fig:flowchart}) is conducted using an in-house built accelerated magnet simulator, ``MagCraft''
({``MagCraft''} can calculate 20,000 observation points for 1,200 magnets within 20 seconds with accuracy. It is detailed in the confidential supplementary document in\,\cite{supplement}).
Subsequently, fitness values are calculated based on the fitness function formulated according to the design objectives, e.g., field strength and field homogeneity etc., and the constraints passed from Step\,1. The fitness function is denoted using $F_\text{base/shim/subgrad}$ or $F_\text{main}$, and the fitness value is denoted as $f^{i,\,j}_\text{base/shim/subgrad}$ or $f^{i}_\text{main}$ for the i$^\text{th}$ individual at the j$^\text{th}$ iteration. The constraints penalize the individual by adding a large value to the fitness value if a condition is not satisfied.  
With the fitness values for the individuals in a population, a stop criterion is applied to check the values to decide whether to stop GA at the current iteration. 
If the stop criterion is met, then the individual with the smallest fitness value is output as the optimized solution. Otherwise, the fitness values of all the individuals are compared, and the good ones are selected for offspring production as the next generation for the next GA iteration where mutation and crossover are applied. The stop criterion here is the relative improvement of the best fitness value: if the average change the best fitness value is less than $10^{-4}$ for 50 generations, the optimization stops.

The details of the GA optimization of each sub-array, including the number of individuals for a population, the defined array of variables for an individual, the fitness function, the applications of constraints, and the stop criteria, are presented in the following sub-sections.
Matlab was used for the implementation. Besides Step\,2, GA was used at the design synthesizer in Step\,1 to facilitate decisions on the range of a parameter.

\subsection{Design \& Optimization of the Main Array}
\label{subsec:design_op_main}
The role of the main array is to supply a strong and homogeneous field (i.e., less inhomogeneous field) . It is aimed to maximize the field strength, and at the same time, minimize the field inhomogeniety to lower the gradient contribution from the main array.
The main array consists of the base array and a shimming array as shown in Fig.\,\ref{fig:main_aray}. As shown in Fig.\,\ref{fig:main_aray}, the base array has a single IO type ring pair with $n_\text{bar,\,base}$ magnet blocks around each ring whereas the shimming array has $m_\text{ring,\,shim}$ rings (i.e., $m_\text{ring,\,shim}$/2 ring pairs), and $n_\text{bar,\,shim}$ magnet blocks around each ring. 
The design and optimization of the main array is shown in the red block in Fig.\,\ref{fig:flowchart}. It consists of the step on physics-guided design and another one on GA optimization. Next, the details of these two steps are presented.

\input{fig_main_array.tex}

\subsubsection{Physics-guided Design of the Main Array}
\label{subsubsec:main_init_design}

Fig.\,\ref{fig:predesign}\,(a) shows the details of the physics-guided design for the main array.
The pre-design inspections and basic understanding for the choices of the design parameters for the base and the shimming arrays, $\Bar{\psi}'_\text{base}$ and $\Bar{\psi}'_\text{shim}$, are obtained separately, which are shown in the two blocks on the left in Fig.\,\ref{fig:predesign}\,(a). Given the availability of magnets off the shelf and the design constraints, an initial range of each parameter can be determined. 

For the base array, the dimension of each block was set to $a_\text{base}\,\times\,b_\text{base}\,\times\,h_\text{base}$\,=\,120\,$\times$\,30\,$\times$\,90\,mm$^3$ based on the availability of the magnet blocks that can be bought off the shelf. The grade of all Neodymium-iron-boron (NdFeB) magnets for the base array was set to N52 with a remanence of $B_\text{r}$= 1.43 \si{\tesla} to maximize the resulting magnetic field strength. $Z'_\text{base}$ is the distance between the inner edges of the inward and outward ring along the $z$-direction as shown in Fig.\,\ref{fig:main_aray}\,(c). Its range was set to $[150,200]$\,mm and that of $R'_\text{base}$ was set to $[160,200]$\,mm, which are the initial settings considering to allow other magnet sub-arrays and hardware, and a cylindrical FoV with a diameter of 20\,cm and a length of 4.5\,cm hosted inside the 
bore. The number of base magnet blocks around the ring, $n_\text{bar,base}$, was varied between 8 and 22, and it was decided to be $n_\text{bar,base}=22$ after a parametric sweep which is detailed in Appendix\,\RomanNumeralCaps{1}.1, considering the sparsity of the array and the resulting field strength and field pattern. 
As $n_\text{bar,base}=22$ was fixed, the radius $R'_\text{base}$ was determined to be the smallest that is possible so that the magnets can be as close to the FoV as possible to obtain a high field strength. With the consideration of housing the magnets, the smallest possible value of $R'_\text{base}$ is 169.55\,mm.
For the polarization of the magnets, the front and the rear ring have radially outward and and inwards polarization, respectively, as shown in Fig.\,\ref{fig:main_aray}\,(a). This is to obtain a magnetic field in the $z$-direction in the FoV, 
as introduced previously.

\input{fig_predesign.tex}

For the shimming array, a ring pair is the basic optimization unit. The choices of the grades of magnets were set to be N45, N48, N50, and N52. An ``air block'' with remanence value of 0 was added as an option to increase the optimization flexibility. Therefore, $B_\text{r}^{\text{shim},\,k}\in$\,[1.32, 1.38, 1.40, 1.43, 0]\,T, the superscript ``$k$'' is the index of the ring pairs. The size of each magnet block was $10\times10\times10\,\text{mm}^3$. It is much smaller than that of the block for the base array, which is the result of two design considerations. One is that, in the axial direction, a small dimension allows more ring pairs which leads to an higher degree of freedom for an optimization for effective shimming. The other is that, in the radial direction, a small dimension occupy less space, thus to reserve space for other hardware (e.g., RF coils). 
A ring pair in the shimming array can be a IO-type or a parallel-type in either $+z$\,- or $-z$\,-\,direction. This leads to $p^k_\text{ring,\,shim}\in\{\pm\text{IO},\,\pm\text{par}\}$.
The range for $R'_{\text{shim}}$ is $[114,138]$ mm set according to the targeted FoV, and the space constraint imposed by the lower bound of $Z'_\text{base}$ at 150\,mm. 
For the number of shimming magnet blocks around the ring, it varies between 8 and 24, $n_\text{bar,shim}$\,=\,[8, 12, 16, 20, 22, 24].
The distance between the inner edges of the two rings in a ring pair, $Z'_\text{shim}$, as shown in Fig.\,\,\ref{fig:main_aray}\,(c), was swept to understand its effects on the field pattern it supplies, thus helping to decide the ranges of the relevant parameters. With this parametric sweep (documented in Appendix \RomanNumeralCaps{1}.2), it is learned that rings with different $Z'_\text{shim}$s offer various concentric field patterns, the higher the number of magnet rings, $m_\text{ring,\,shim}$, the higher the degree of freedom for optimization. On the other hand, the axial length of the shimming array cannot be longer than that of the base array to facilitate the access to the bore, i.e., ($a'_\text{shim}$\,$\times$\,$m_\text{ring,\,shim})\,\le$\,($Z'_\text{base}+2h'_\text{base}$). Meanwhile, as shown in Appendix \RomanNumeralCaps{1}.2, larger the $Z'_\text{shim}$ when the rings are further apart, the fields in the FoV is smaller, and thus less contribution it has to the total fields. Therefore, $m_\text{ring,\,shim}$ was set to 38 where the end-to-end length of the shimming array is the same as that of the base array when $Z'_\text{base}$ is set to the maximum, 200\,mm.

Following the pre-design inspections, as shown in Fig.\,\ref{fig:predesign}\,(a), the output design parameters, $\Bar{\psi}'_\text{base}$ and $\Bar{\psi}'_\text{shim}$, are input to a design synthesizer. The aim of this design synthesizer is combining  the base and shimming arrays to further tailor the ranges of the design parameters to achieve the targets, a high field strength and high field homogeneity. For a high field strength, a minimum constraint for the main field strength is set, i.e., mean($B_0$)\,=\,mean($B_\text{base}$\,+\,$B_\text{shim}$)\,$>$\,100\,mT. 
For a high field homogeneity,   
the effect of the number of magnet bars around a ring of the shimming array on the $B_0$ homogeneity was checked.
Based on a comparison among the GA optimized $B_0$s at $n_\text{bar,shim}$\,=\, 8, 12, 16, 20, 22, 24 with $m_\text{ring,\,shim}$\,=\,38, $n_\text{bar,\,base}$\,=\,22, $Z'_\text{base}$\,=\,200\,mm (detailed in Appendix \RomanNumeralCaps{1}.2), the higher the $n_\text{bar,shim}$, the higher the total field homogeneity. Therefore, $n_\text{bar,shim}$ was decided to be 24, balancing the field homogeneity and the weight of the array.
Meanwhile, since both the base and the shimming arrays provide concentric field pattern, a straightforward way to obtain a homogeneous $B_0$ is to keep the difference and the linearity of their fields close. 
Two inspections were conducted, one on the field difference, $\Delta B_\text{base}$, and the linearity when $Z'_\text{base}$ varies (detailed in Appendix \RomanNumeralCaps{1}.1), and another on the matching of $\Delta B$ through a GA optimization with and without a constraint on the linearity at $m_\text{ring,\,shim}$\,=\,38 (detailed in Appendix \RomanNumeralCaps{1}.3). Based on these inspections, the lower bound of $Z'_{base}$ is refined to be 180\,mm for the next step of optimization.
A summary of the finalized ranges of the design parameters, $\Bar{\psi}_\text{main}$, design constraints, and preset values are listed in the right most block in Fig.\,\ref{fig:predesign}\,(a). They are input to the GA step.
It should be noted that the design synthesizer is dynamic. There is a feedback loop from the checking on the performance of the optimized magnetic field on image reconstruction, as shown in Fig.\,\ref{fig:flowchart} and Fig.\,\ref{fig:predesign}\,(a). 
If the image quality does not meet the requirement, the constraint on minimum $B_0$ needs to be changed in the design synthesizer in Fig.\,\ref{fig:predesign}\,(a).

\subsubsection{GA Optimization of the Main Array}
\label{subsubsec:main_op}

The field of the base array and that of the shimming array are combined for GA optimization. The inputs to the GA block are from the previous step. 
Therefore, with $m_\text{ring,\,shim}$\,=\,38 (i.e., 19 ring pairs) for the shimming array, 
the variables for the optimization of the main array are included in the array, $\Bar{\psi}_\text{main}$\,=\,$[B^\text{shim,\,1}_r;\,B^\text{shim,\,2}_r;\,\dots;\,B^\text{shim,\,19}_r;\,p^1_\text{shim};\,p^2_\text{shim};\,\dots;\,p^{19}_\text{shim};\\ \,R'_\text{shim};\,Z''_\text{base}]$ where each variable have the ranges of values decided at the physics-guided design as shown in the box on the right in Fig.\,\ref{fig:predesign}\,(a).

For the GA for the design of the main array, the fitness function is expressed by,
\begin{equation}
\label{eq:fitness_function1}
F_\text{main}=\argmin_{\Bar{\bm{\psi}}_\text{main}} \bigg\{\Delta B_{0,1D}+P_1(B_0)\bigg\}
\end{equation}
where $B_{0,1D}$ is the 1D main magnetic field distribution along the $+\,x$-axis and $P_1(B_{0,1D})$ is the penalty for the constraint, $B_0\,\ge\,$100\,mT. If $\text{mean}(B_{0,1D})<100$ mT, $P_1(B_{0,1D})=1000$, and it equals to 0 otherwise. 
In this fitness function, only the 1D field in the radial direction is used instead of the whole 2D maps because the main field is axially symmetric, and the 1D field distribution can provide enough information for the optimization while requiring less computational power. The number of individuals in each population was set to 50 and 100.

\subsection{Design \& Optimization of the Sub-Gradient Array}
\label{subsec:design_op_grad}
The role of the sub-gradient array is to supply a gradient field that has ``good encoding capability'' when it is rotated for imaging (i.e. leads to good quality image)
yet has an inhomogeneity that is within the bandwidth of the RF system. As a rotation of the gradient field can become necessary for encoding and imaging, the array is desired to be light and has a small tangential magnetic force to lower the load of a rotation mechanism, so as to increase the rotation accuracy.
Therefore, the optimization objectives of the design of the sub-gradient array are set to have a monotonic field pattern within a field homogeneity that RF coils can tolerate (0.1$\times$mean($B_0$) for an RF bandwidth of no more than 10\%), and to have a light weight. A double ring array with an offset between the rings, named offset-double-ring array, was designed and optimized to approach these objectives.

\input{fig_ring_field_current_model.tex}
\input{fig_subgradient_array.tex}

\subsubsection{The design of an offset-double-ring array}
\label{subsubsec:subgrad_design}

When a ring magnet has an inner radius, $R$, that is much greater than the thickness of the wall of the ring, $\Delta R$, i.e. $R\gg\Delta R$, and the length of the ring, $\Delta Z$ is much greater than $\Delta R$ as well, it can be used to design an array that supplies a field with a monotonic pattern. Fig.\,\ref{fig:ring_magnet}\,(a) and (b) show the 3D view and front view of such a ring magnet that has its axis aligned with the z-axis, $R\gg\Delta R$, and $\Delta Z\gg \Delta R$, and the distribution of the z-component of the magnetic field supplied by the magnet within a circle with a diameter of 200\,mm on $xy$-plane when the center of the ring magnet is at the origin.
As shown, this field pattern is concentric. Therefore, mathematically it can be expressed as $B_\text{z} (x,y)=\sum^N_{n=0}\;C_n\;(x^{2n}+y^{2n})$, where $C_n$ is the coefficient of the $n^\text{th}$ order term.
When the magnet ring has a large $R$ and $\Delta Z$, and a small $\Delta R$, the higher order terms can be neglected, and the expression of $B_\text{z} (x,y)$ can be simplified as,
\begin{equation}
\label{eq:2nd_order_appro}
B_\text{z}(x, y) = C_0+C_1(x^2 + y^2)
\end{equation}
For the detailed derivation, please refer to Appendix \RomanNumeralCaps{2}.

When Eq.\,(\ref{eq:2nd_order_appro}) holds, an offset of the ring by $\Delta x$ in the $x$-direction results in a field pattern that can be described by the following expression,  $B_\text{z}((x-\Delta x), y) = C_0+C_1((x-\Delta x)^2 + y^2)$. Therefore, when two identical rings, one has an offset of $\Delta x_1$ along the $-x$-direction and supplies a field pointing in the $+z$-direction (called $+z$-polarized ring and colored in blue as shown in Fig.\,\ref{fig:ring_magnet}\,(c)), and the other one has an offset of $\Delta x_2$ along the $+x$-direction and supplies a field pointing in the $-z$-direction (called $-z$-polarized ring and colored in green as shown in Fig.\,\ref{fig:ring_magnet}\,(d)), are combined, their fields are superposed, i.e., the field of the former magnet is subtracted by that of the latter one, resulting in a field that is expressed as $B_\text{z}(x, y) = 2x(\Delta x_1 - \Delta x_2)$, which has a linear gradient along the $x$-direction, as illustrated in Fig.\,\ref{fig:ring_magnet}\,(e).

For a magnet ring that has $R$\,=\,230\,mm, $\Delta R$\,=\,20\,mm, and $\Delta Z$\,=\,200\,mm, the field pattern can be modeled using Eq.\,(\ref{eq:2nd_order_appro}). Fig.\,\ref{fig:ring_magnet}\,(c) and (d) shows the field pattern (in the $+z$-direction) of the $+z$-polarized ring with an offset of $\Delta x$\,=\,8\,mm in the $-x$-direction and that of the $-z$-polarized ring (in the $-z$-direction) with the same amount of offset in the $+x$-direction, respectively. Fig.\,\ref{fig:ring_magnet}\,(e) shows the field pattern when these two offset rings are combined, i.e., an offset-double-ring array, which is linear. 
Furthermore, the length of the array in the $z$-direction affects the field pattern.
When $\Delta Z$ decreases, the higher-order terms in Eq.\,(\ref{eq:2nd_order_appro}) need to be considered, the linearity of the pattern is compromised. 
Fig.\,\ref{fig:ring_magnet}\,(f) shows the field pattern when the $\Delta Z$ is reduced to 10\,mm. Comparing Fig.\,\ref{fig:ring_magnet}\,(f) to Fig.\,\ref{fig:ring_magnet}\,(e), the field is curved, in other words, the field pattern deviated from a linear pattern when $\Delta Z$ decreases.

For a practical implementation, each magnet ring of the offset-double-ring array was discretized into $n_\text{bar}$ magnet bars, and each bar is discretized into $N_g$ blocks, i.e., each ring is discretized into $N_g$ short rings, as shown in Fig.\,\ref{fig:subgradient_array}. Fig.\,\ref{fig:subgradient_array}\,(b) and (c) show the side view and the cross-sectional view of the sub-gradient array, respectively. 
Besides being easy to implement, By doing this, the magnet volume of the array can be significantly reduced, which reduces the weight of the array and the force it experiences. 

As a result of the discretization, the sub-gradient array becomes sparse, light, and easy to assemble. However, the monotonicity of the field pattern is compromised. On the other hand, the inhomogeneity of the field need to be controlled within the defined design range. Therefore,  optimizations are needed to obtain the structural sparsity, field monotonicity and homogeneity at the same time.

\subsubsection{The physics-guided design of sub-gradient array}
\label{subsubsec:subgrad_init_design}

Fig.\,\ref{fig:predesign}\,(b) shows the physics-guided design for the sub-gradient array. The pre-design inspections were conducted for the choices of the design parameters, $\Bar{\psi}'_\text{subgrad}$. The details are shown in the block on the left of Fig.\,\ref{fig:predesign}\,(b). As shown, based on the availability of magnets, the grade of magnets for sub-gradient array was chosen from N45, N48, N50 and N52 NdFeB magnets with remanence of $B_\text{r}=1.32, 1.38, 1.40, \text{ and } 1.43$ \si{\tesla}, respectively.   
For the setting of other parameters, 
the dimension of each magnet block was set to $a'_\text{g}\,{\times}\,b'_\text{g}\,{\times}\,h'_\text{g}=$ 20\,$\times$\,20\,$\times$\,20\,mm$^3$ based on the availability of the magnet blocks and the need for high flexibility in the sub-gradient array optimization. The inner radius of each magnet column, $R'_{\text{g}i}$, is set separately where $i$ indexes the columns of the whole offset-double-ring, as shown in Fig.\,\ref{fig:subgradient_array}\,(c). The range of $R'_{\text{g}i}$ was set to $[210,240]$\,mm, which falls in between $R'_\text{base}$ and $R'_\text{shim}$. The off-center distance of the sub-gradient rings, $\Delta x$, was chosen within $[5,15]$\,mm, limited by the physical confinement of the base array. The number of magnet columns in each sub-gradient ring had $n_\text{bar}^\text{p/n}\in [4,36]$, and the number of magnets in each column $N_\text{g}\in [2,18]$ for high-flexibility optimization. 

Similar to the main array, the output of the pre-design inspections, $\Bar{\psi}'_\text{subgrad}$ as well as the optimized main $B_0$ field, are input to the design synthesizer, as shown in the middle block of Fig.\,\ref{fig:predesign}\,(b). 
The aim of the design synthesizer here is to further tailor the ranges of the design parameters, by combining the gradient fields ($G$) and the optimized $B_0$,  to approach the targets, linearity of the total field ($G+B_0$), and a well-controlled approximate gradient $\Delta G/\Delta x$.
A big $N_\text{g}$ or a big $n_\text{bar}^\text{p/n}$ lead to higher $\Delta G$ and more linearity, which indicates better encoding capability for imaging. However, they result in an increase in the weight of the magnet array.
To balance the encoding capability with the weight of the sub-gradient array and to keep $\Delta G$ within the targeted bandwidth, the number of rings and magnet bars were preset to be $N_\text{g}=4$ and $n_\text{bar}^\text{p/n}=10$.
Also, a target gradient $\nabla G_\text{ref}$ is defined for the control of $\Delta G/\Delta x$. During GA optimization, only the designs with $\Delta G/\Delta x$ having less than 5\% deviation from $\nabla G_\text{ref}$ will be regarded as valid.
A summary of the finalized ranges of the design parameters, $\Bar{\psi}_\text{subgrad}$, design constraints, and preset values are listed in the right most block in Fig.\,\ref{fig:predesign}\,(b). They are input to the GA step.
Similarly to the step for the main array, the design synthesizer is dynamic. There is a feedback loop from the checking on the performance of the optimized magnetic field on image reconstruction, as shown in Fig.\,\ref{fig:flowchart} and Fig.\,\ref{fig:predesign}\,(b). If the image quality does not meet the requirement, the target of $\Delta G/\Delta x$ will be set higher within the range of $\nabla G_\text{ref}\in[10,\,40]$ mT/m, which is within the tolerable range.

\subsubsection{GA optimization of sub-gradient array design}
\label{subsubsec:subgrad_op}
GA was used for the optimization. The objectives are a field monotoncity along the $x$-direction, and a field inhomogeneity that is less than 10\% of the average $B_0$ strength within the FoV. Another objective is the sparsity of magnets, i.e., a small number of magnets, for a light weight and a small magnetic force to lower the load of a rotation mechanism so as to reduce the rotation error. The inputs are the design variables, constraints, and preset values from the previous step. 

For the optimization, the variables are the inner radius $R'_{\text{g}i}$, the remanence $B_\text{r}^{\text{g},i}$, and the off-center distance $\Delta x$, forming $\Bar{\bm{\psi}}_{g}=[R'_{\text{g}1}\enspace R'_{\text{g}2}\enspace \dots R'_{\text{g}10}\enspace B_\text{r}^{\text{g},1}\enspace B_\text{r}^{\text{g},2}
\enspace \dots B_\text{r}^{\text{g},10}\enspace \Delta x]$. The proposed accelerated magnet simulator, ``MagCraft''\,\cite{supplement}, was used for the forward calculation of the gradient field, $G$. The fitness function is written as follows,
\begin{equation}
\label{eq:fitness_function2}
F_\text{subgrad}=\argmin_{\Bar{\bm{\psi}}_\text{g}} \bigg\{P_2(G,\nabla G_\text{ref})-R^2_\text{total}\bigg\}
\end{equation}
where $P_2(G,\nabla G_\text{ref})$ is the penalty for the constraint, $0.95\nabla G_\text{ref}<\Delta G/\Delta x<1.05\nabla G_\text{ref}$. If $\Delta G/\Delta x$ does not fall into this range, $P_2(G,\nabla G_\text{ref})=1000$ and it equals to 0 otherwise. 
And $R^2_\text{total}$ is the linear regression coefficient of the total magnetic field $B_0+G$, indicating the goodness of the linear surface fitting. In the optimization, both $R_{\text{g}i}$ and $\Delta x$ takes a step size of 1\,mm.  The stop criteria were set to end the optimization upon the saturation of the fitness value (with a tolerance of $10^{-4}$) for 50 generations. These settings are the same to those of the main array optimization. After the optimization, $[R'_{\text{g}1}\enspace R'_{\text{g}2}\enspace \dots R'_{\text{g}10}]$ are checked via a 3D modelling of the structure to avoid overlapping of the magnet columns.  
The population size was set to 50 and 100, which is the same as the optimization of the main array.

\input{fig_field_3D.tex}
\input{tab_result_3D.tex}

\subsection{Encoding Capability Checking for ($B_0+G$)}
\label{subsec:encode_capability}
After each round of GA optimization for the PMA design, image reconstruction is performed to check the encoding capability of the resulting magnetic field pattern, as shown at the bottom part of Fig.\,\ref{fig:flowchart}. ($B_0+G$) was rotated to encode signals and the Shepp-Logan phantom was used. The structural similarity (SSIM) index of the reconstructed image was used to check the image quality and aliasing was examined. 
When the SSIM of the latter case is less than the former ones, it is considered as unsatisfactory, and the design synthesizer for the red and/or the blue blocks is re-triggered.

The aim for the encoding capability check is to produce a PMA design that gives the SSIM of reconstructed image as high as possible. Once the SSIM cannot have an increase that is larger than 0.001, the re-triggering of the design synthesizer is stopped and the optimized results are output.

\section{Results}
\label{sec:results}

\subsection{Magnetic Field Distribution of the Proposed PMA}
\label{subsec:result_field}

Fig.\,\ref{fig:field_3D} shows the field distributions in the $z$-direction of the slices at $z=0$, 5, 10, 15 and 20\,mm within the targeted FoV.
The slices are the center slice of each 5\,mm-thick cylinder as indicated in dark grey in Fig.\,\ref{fig:3D_overview}. Due to the symmetry with respect to the $xy$-plane, only the fields of the slices on the rights are shown and analyzed.
Table\,\ref{tab:result_3D} lists the characteristics of the fields of each slice. 
As shown in Fig.\,\ref{fig:field_3D}, from the center slice towards the end in the FoV, the fields show a off-concentric pattern with a circular center bright zone with increasing intensity. Thus, it has gradient in the $y$-direction. These patterns are approximately monotonic, which can work with a linear gradient coils in the other direction for Fourier imaging with corrections\,\cite{cooley2021Nature_BE}. They can be rotated for signal encoding for imaging. They encoding capabilities will be further examined in the next subsection, Section\,\ref{subsec:result_recon}. The change of the field pattern at difference slices is accompanied with a change in the mean of the field strength and the field difference, $\Delta(B_0+G)$.

For the characteristics of the fields supplied by the proposed PMA at different slices, more specifically, as shown in Table\,\ref{tab:result_3D}, the mean of the total field is in the range of 111.40\,mT to 110.79\,mT with a change of 0.55\% moving from the center slice to the edge of the FoV. For the field differences, it is kept within 11\,mT in the FoV, corresponding to an RF bandwidth of $<$\,10\%. It varies from 5.94\,mT at the center slice (a gradient of 29.70\,mT/m) to 10.57\,mT at the slice at the edge (a gradient of 52.85\,mT/m). Therefore, a slice selection can be done by applying a $z$-gradient and the required bandwidth of the RF coil is 10\%.

The optimized parameters are, $R'_\text{base}$\,=\,169.55\,mm, $Z'_\text{base}$\,=\,193\,mm, $R'_\text{shim}$\,=\,121\,mm, and $\Delta x$\,=\,15\,mm. The shimming array is optimized with optimal polarization and remanence of each ring pair shown by the arrows and colors in Fig.\,\ref{fig:main_aray}\,(c), respectively. For the sub-gradient array, the optimal remanence and $R_{\text{g}i}$ of each magnet column are shown in Table\,\ref{tab:subgrad_detail}.

\input{tab_subgrad_detail.tex}

\input{fig_field_map_total.tex}

\input{fig_field_maps2.tex}

Fig.\,\ref{fig:field_map_total}\,(a) and (b) show the $x$-, and $y$-components of the total magnetic field of the center slice at $z=0$\,mm of the proposed PMA. Comparing to the $z$-component of the field of the same slice shown in Fig.\,\ref{fig:field_3D}\,(a), the field in the $z$-direction is much higher than those in the other two directions. In other words, the fields supplied by the proposed PMA is mainly in the $z$-direction, and those in the other directions are negligible.

The proposed PMA consists of the main array and the sub-gradient one. The fields of different sub-arrays are compared using those of the center slice. 
Fig.\,\ref{fig:field_map2}\,(a)-(d) show the fields of the sub-arrays, and  Fig.\,\ref{fig:field_map2}\,(e) shows the corresponding 1D fields along the $x$-axis as well as that of the total field in Fig.\,\ref{fig:field_map2}, all on the center slice.
The field of the main array, $\textbf{B}_0$, is shown in Fig.\,\ref{fig:field_map2}\,(a) with the same color scale as that in Fig.\,\ref{fig:field_map_total}\,(a).
As shown in Fig.\,\ref{fig:field_map2}\,(a), the main array is optimized to supply a homogeneous magnetic field with an average field strength of \SI{111.40}{\milli\tesla} and an inhomogeneity of \SI{1.27}{\milli\tesla} (11400\,ppm/1.14\%/\SI{54.1}{\kilo\hertz}). The high field strength is contributed by the base array and the homogeneity is owing to the effective shimming by the optimized shimming array. 

Fig.\,\ref{fig:field_map2}\,(b) and (c) show the fields of these two sub-arrays. For the base array, as shown in Fig.\,\ref{fig:field_map2}\,(b) and by the black dotted line in Fig.\,\ref{fig:field_map2}\,(e), the base field is high near the circumference, and decays inward towards the center. 
The inhomogeneity of 14.70\,mT is large which on one hand, consequently overshadows the pattern of the sub-gradient array, and on the other hand, goes far beyond the tolerable RF bandwidth. Therefore, a shimming array was designed to reduce the inhomogeneity.
The field distribution of the shimming array is shown in Fig.\,\ref{fig:field_map2}\,(c) and by the red dotted line in Fig.\,\ref{fig:field_map2}\,(e). As can be seen, the fields supplied by the shimming array are negative (i.e., in the $-z$-direction) where it is more negative in the circumferential regions and less negative in the central region. When this field is added to that of the base array (Fig.\,\ref{fig:field_map2}\,(c)), a field that has high homogeneity is successfully obtained as shown in Fig.\,\ref{fig:field_map2}\,(a).  
It is an improvement in homogeneity by nearly 10 times compared to that of a base array only (Fig.\,\ref{fig:field_map2}\,(b)) without a shimming array. The shimming effect can be seen by comparing the 1D field distributions of the main array (after shimming) and that of the base array (before shimming) shown by the black dashed and dotted lines, respectively, in Fig.\,\ref{fig:field_map2}\,(e). The down side of the shimming is a reduction in the field strength as the shimming array was designed to point in an opposite direction for a field subtraction from that of the base array, which is due to the decaying nature of the fields of permanent magnets. In other words, with the nature of the fields of permanent magnets, a shimming field which both shims and increases/aligns with the field of the base array is not possible.

For the sub-gradient array, it supplies a monotonic pattern with a zero average strength, an inhomogeneity of 5.91\si{\milli\tesla}, and a gradient of about 25.99\,mT/m, as shown in Fig.\,\ref{fig:field_map2}\,(d). The corresponding 1D plot along the $x$-axis is shown by the red dashed curve in Fig.\,\ref{fig:field_map2}\,(f). 
The inhomogeneity corresponds to an RF frequency bandwidth of \SI{251.6}{\kilo\hertz}.

Comparing the main array and the sub-gradient array, the former has a much higher average field strength and a much higher homogeneity compared to the later. In this proposed magnet array, the field strength is mainly contributed by the main array whereas the field pattern is mainly contributed by the sub-gradient array.

\subsection{Evaluation of the encoding capability}
\label{subsec:result_recon}

\input{fig_simulated_images_3D.tex}

\input{fig_simulated_images.tex}
For the fields supplied by the proposed PMA, as observed in the previous subsection, when the slice is move from the center to the end of the FoV, the field pattern has the brighter central zone, the average field strength decreases by 0.55\%, and the field difference increases from 5.94\,mT to 10.57\,mT. 
When the fields are rotated for encoding for imaging, they show different the encoding capability which was examined by checking the simulated reconstructed images using conjugate gradient for image reconstruction and a Shepp-Logan phantom. Fig.\,\ref{fig:simulated_images_3D} shows the reconstructed images at different number of rotation angles (denoted as $N$) at Column\,2-4, and the corresponding local $k$-spaces when the FoV is split into $7\times 7$ sub-FoVs when the number of rotation angles is 144. The structural similarity (SSIM) index of each image was calculated and shown in red on the top of each image. 
For all the images, the resolutions are 100\,$\times$\,100 in 1\,mm\,$\times$\,1\,mm. 
The maximum rotation angles were set to guarantee the best image quality in each case\,\cite{jia2019effects}, which is 360$^{\circ}$ for all the three cases. For the reconstruction, the coil sensitivity was set to be uniform and the SNR was set to 15\,dB to mimic the noise level in a low-field system. Local $k$-spaces can be used to explain the relation of imaging quality and the encoding field\,\cite{jia2019effects}. 

As shown at the first row in Fig.\,\ref{fig:simulated_images_3D}, at the center slice, the reconstructed image can reach a SSIM of 0.836. For the local $k$-space, each sub-FoV has a reasonable filling factor which indicates high gradients in all the regions although the gradients are not linear, and leads to good encoding and subsequent high-quality image. Meanwhile, it is noticed that, based on the distributions of the points, the gradients along the northeast direction is smaller than that along the northwest one. When it is moved to the next 5-mm slice, as shown at the second row in Fig.\,\ref{fig:simulated_images_3D}, similar performance and local $k$-space are observed. 
For the next three 5-mm slices, the reconstructed images degrades considerably and artifacts are seen, which is due to a gradient contrast in different directions in the peripheral regions. The gradient contrast can be seen in the peripheral sub-FoVs of the local $k$-spaces. At Row\,3, the peripheral local $k$-spaces become tilted 8-shapes as the gradients in the northeast direction approaches zero and those in the northwest direction are large. At Row\,4, the gradient contrast of the slice is larger where the peripheral local $k$-spaces becomes a dot as the gradients in the northeast direction approaches zero and those in the northwest direction are further increased and the points are out of the plotting range. 
At Row\,5, the local $k$-space pattern has lower gradient contrast than that of Row\,4. As for the image quality, although the SSIM values of Row\,5 are lower than those of Row\,4, it can be observed that Row\,5 has better contrast and less salt-and-pepper noise over the phantom. This finding suggests that the SSIM value cannot reflect all aspects of image quality.

The encoding capability of the field pattern supplied by the proposed PMA is compared to a linear pattern and one of a short Halbach array\,\cite{ren2015magnet}.
For this comparison, 
the patterns of the center slices at $z=0$\,mm in the PMAs are examined. 
Fig.\,\ref{fig:simulated_images} shows the field patterns (Column\,1), reconstructed images (Column\,2-4 for $N=36$, 72, 144, respectively) and the corresponding local $k$-spaces (Column\,5, 7\,$\times$\,7 sub-FoVs, at $N=144$). 
For a fair comparison, they all have an average field strength of 111.40\,mT and a field inhomogeneity of 5.91\,mT. The linear field pattern has a gradient of 29.55\,mT/m. 
Comparing the reconstructed images in Fig.\,\ref{fig:simulated_images}, as can be seen, at $N$\,=\,36, the image using the field from the proposed array has much better image quality compared to the linear pattern and the Halbach case. The linear case has some artifacts due to the lack of gradient in the $y$-direction. Moreover, the Halbach case has an obvious central blurry region which is due to the low/zero gradient in that region. 
When the number of rotation angles increases, the imaging quality in both the proposed PMA and the linear cases improve considerably, and it is noticeable that the former is always better than the latter. At $N$\,=\,72, the proposed PMA case has a higher SSIM by 0.089 compared to the linear case, and at $N$\,=\,144, the former is higher by 0.125. 
For the Halbach case, it has an improvement in image quality when the number of the rotation angles increases, but the improvement is limited, and is much lower than the other two cases. 

For the local $k$-spaces at Column\,5 in Fig.\,\ref{fig:simulated_images}, at $N$\,=\,144, the field of the proposed PMA and the linear one have similar coverage of signal points at the central sub-regions, and the proposed PMA has more coverage at the peripheral ones in either the northeast or the northwest directions. This is owing to the gradient in the y-direction. It is the reason that the proposed array leads to a higher image quality when it is used for encoding. For the Halbach case, due to the low/zero gradient fields at the center of the FoV, its local $k$-space in the central sub-FoV is shrink to a dot, which leads to the central blurry region in the image.
When comparing the Halbach case to the proposed case, the later has a better coverage in all the sub-FoVs, including the central one. 
In the central sub-FoV in the Halbach case, as the gradient is low or approaching zero, the coverage of the signal points in the local $k$-space does not improve when the number of rotation angles increases. It is the reason that in the reconstructed image, this region stays blurry even when $N$ increases. In the peripheral region, the Halbach local $k$-space tends to be thinner and even becomes a line, which dramatically reduced the coverage. Thus, when $N$ increases, SSIM of the Halbach case does not increase much, and is much lower than the other two cases.

\input{fig_PSF.tex}

The alias in the simulated images in Fig.\,\ref{fig:simulated_images} when different field patterns are used for encoding are analyzed using point spread function (PSF).
Fig.\,\ref{fig:fig_PSF} shows the images with the PSF's of two pixels, one at the center at (0,0)\,[mm] and the other at the peripheral region of the phantom at (0,90)\,[mm] in an MRI system at $N=36$ using the field pattern supplied by the proposed PMA, a linear pattern and a Halbach pattern shown at Column\,1 in Fig.\,\ref{fig:simulated_images}. The coil sensitivity was set to be uniform. 
As shown at Row\,1 in Fig.\,\ref{fig:fig_PSF}, with the rotation of the field, the center pixel of the pattern of the proposed PMA and that of the linear pattern contaminate the rest of the image in a similar radially pattern except the former shows distortion of the rays and the latter shows straight lines with an increased intensity. For these two patterns, similar PSF plots are observed from the pixel at the peripheral region, and comparing the two, they show similar difference. For the short Halbach array, as shown at Column\,3 in Fig.\,\ref{fig:fig_PSF}, the center pixel contaminates both the central and the peripheral regions whereas the peripheral one contaminates the whole FoV unevenly.  

The encoding capability of the proposed PMA is examined when it is rotated for imaging. The three slices are the center can lead to reconstructed images with SSIM of more than 0.83 when signals are acquired at 144 angles. The other six slices toward the ends of the FoV shows compromised encoding capabilities (with an SSIM between 0.511 to 0.638 for the simulated reconstructed images) as the field difference becomes larger, the gradient contrast goes higher, and the gradient becomes very small in some directions in the peripheral sub-FoVs. The proposed PMA can lead to high quality imaging with a slice selection. 
The field pattern of the proposed PMA is further compared to a linear patter and one from a short Halbach array. Based on the comparison, the proposed one offers higher gradient in some directions in the peripheral sub-FOVs without additional alias due to imperfection in linearity, which leads to the best encoding capability among the three under comparison. 
Compare to the field from a Halbach array, the proposed PMA does not have a region where the gradient is very low or approaching zero. Moreover, its field is not symmetric so that an extra encoding using coil sensitivity may not be needed.

\section{Discussions}
\label{sec:discussions}

The proposed IO ring-pair array has the highest average field strength at 111.40\,mT for head imaging, and the highest magnetic field generation efficiency at 0.88\,mT/kg among all the PMAs reported in the literature. Permanent magnets generate fields that decay at the order of 1/$r^3$. Higher the volume of the magnet leads to higher the magnetic field\,\cite{Furlani2001book}. Thus, more magnets in an array help to have a higher field strength in the FoV. Magnetic field generation efficiency, that is calculated by using the average field strength in the targeted FoV divided by the total weight of the array, can be a parameter that tells how efficient a PMA design is. The highest magnetic field generation efficiency of the proposed PMA here indicates that the array has the most efficient arrangement of the magnets in terms of the choice of the magnet grades and size, the magnet grouping, and the locations and orientations of the magnets/magnet columns.

In the proposed PMA, the direction of the field is in the axial direction of the cylinder, which is the same as a conventional superconducting-magnet-based MRI system. It allows the application of surface loop coils to the system without compromising the field efficiency when the magnet rotate. Meanwhile, it guarantees the same efficiency when loops are populated around the wall of the cylinder for multi-channel imaging. Moreover, this allows the applications of other high-performance RF coils that were designed for a transitional MRI system, and the applications of other advanced MR techniques. It opens more opportunities of the combinations of the technological advancements developed for a traditional system and a PMA system. 

For the inhomogeniety, the proposed PMA has the field inhomogeneity controlled within 11\,mT for the whole FoV, which corresponds to a RF bandwidth of less than 10\%. A commercially available spectrometer can withstand a 10\% bandwidth, whereas 10\%-bandwidth RF coils that have transversal $B_1$-fields are designable\,\cite{tommy2012coil}.  

The proposed PMA shows an unique off-concentric pattern with a circular center bright zone at different slices in the targeted FoV. It is approximately monotonic, thus it can work with linear gradient coils in the other two directions for Fourier imaging with corrections\,\cite{cooley2021Nature_BE}. When it is rotated to encode signals alone for imaging, the unique pattern with the gradient in the y-direction offers better encoding capability and produce images with better qualities compared to a linear pattern. A similar comparison and result are reported in\,\cite{gong2019local_k}.

As the field of the proposed PMA can be rotated to encode signals for imaging, the magnet array is designed in such a way that the rotation of the field can be realized by rotating a light sub-gradient array. This significantly lowers the burdens of the mechanical system, which helps to guarantee the accuracy of rotation angles and to reduce the power input of the system.


For the GA optimization, SSIM of the reconstructed image using the magnetic field of the proposed array for encoding was not used as a fitness function for the following reasons. One is that it is extremely time-consuming. It takes more than 800 times longer for one iteration compared to one presented in Section\,\ref{sec:methods} above even when the resolution of the image was set to be 50, half of those shown in  Fig.\,\ref{fig:simulated_images} and Fig.\,\ref{fig:simulated_images_3D}. The second reason is that  
the GA optimization result using low image resolution may not guarantee a high SSIM when the optimized PMA is tested with high-resolution image reconstruction. 
Therefore, using SSIM as a fitness function is yet to be practical before the image reconstruction is accelerated. 
Besides, there are other indicators for image quality without performing image reconstruction, such as the local $k$-space pattern, they will be explored to be used as fitness values for PMA optimization in the near future. 

\input{fig_5_Gauss.tex}

The 5-Gauss region is examined. 
Fig.\,\ref{fig:5Gauss} shows the total magnetic fields ($|B|=\sqrt{B_\text{x}^2+B_\text{y}^2+B_\text{z}^2}$) of the proposed PMA in a wider region to check the fringing field.  Fig.\,\ref{fig:5Gauss}\,(a) and (b) plots the 5-Gauss safety lines on $x$-$y$ and $x$-$z$ planes, respectively. The location of the PMA is indicated using red dashed lines. As shown in Fig.\,\ref{fig:5Gauss}\,(a), in either the $x$- or the $y$- direction, the 5-Gauss line is 870\,mm away from the center, while in the $z$-direction as shown in Fig.\,\ref{fig:5Gauss}\,(b), it is at $z=1040$ mm. It can be concluded that the fringing field becomes negligible at a distance of 1\,m from the center of the FoV.

The magnetic force experienced by each magnet block inside the proposed PMA was calculated to choose the housing material. The force calculation was done by using a in-house built code that is included in ``MagCraft'' with a validation included in Appendix\,III.
The maximum force is 231.23\,N on an area of 2700\,mm$^2$, which can be supported by most of housing materials, such as Nylon and aluminium. 
Moreover, the force calculation was applied to calculate the tangential force experienced by the sub-gradient array, to evaluate the force to the load of the mechanical rotation system when a rotation is needed. When the sub-gradient array is in the initial position as shown in Fig.\,\ref{fig:array}, it experiences a force of 6.56\,N and 0.36\,N in the $x$- and the $y$-direction, respectively, which indicates that the magnetic force does not contribute much to the radial force of the mechanical rotation system to add a destructive load. 

Furthermore, to facilitate an assembly, the force calculation was applied to optimize the insertion sequence to minimize the sum of forces experienced by each block of the base array, and each column of the shimming and sub-gradient arrays that is being inserted into the housing structure.
To adapt with the housing design, the general assembly sequence is the following, the first base ring, the sub-gradient array, the second base ring, and the shimming array. After a GA optimization, an optimal sequence of magnet insertion for each sub-array is produced to lower the maximum force experienced by the magnet unit during insertion. An example of such a GA optimization is shown in Appendix\,\RomanNumeralCaps{4}.
Among the calculated forces, the largest force experienced by a single magnet unit, one magnet block for the base array or one magnet column for the shimming and the sub-gradient array, is about 350\,N, which happens 4 times for the blocks of the base array. It should be noted that during the insertion, each sub-gradient column are supposed to be treated as a single magnet, and each shimming column should be assembled before its insertion into the housing of the PMA. For the insertion of the columns of the shimming array, the forces experienced by each unit are below 20\,N. For the insertion of the sub-gradient array, the force experienced by each unit is below 80\,N. The building of the proposed PMA is ongoing. As the focus of the paper is the design of the magnet array, the 
forces experienced by the magnet units during assembly are examined to be practical, together with the accuracy of calculation of magnetic field of magnet array and the accuracy of the fabrication of magnet housing, this paper can be a stand alone worthwhile sharing of know how before the assembly is finished. 

For the temperature drift of the magnets at room temperature, a field characterization was conducted on a 10\,mm N52 cube magnet within a normal temperature fluctuation between \SI{22.9}{\celsius} and \SI{25.7}{\celsius} in the lab. A drift of less than 1\% was observed on a distance of 12\,mm away from the magnet surface. The drift of the field can vary when the range is bigger. In general, the proposed PMA that consists of NdFeB magnet blocks can provide stable magnetic field under a typical range of change at room temperature. The details of the testing are included in Appendix\,\RomanNumeralCaps{5}.

\section{Conclusion}
\label{sec:conclusion}

In this paper, we propose a sparse cylindrical PMA that generates a strong field along the axial direction with a monotonical pattern. 
The magnet supplies an average field strength of 111.40\,mT and an inhomogeneity of 10.57\,mT within a cylindrical FoV of 20\,cm in diameter and 4.5\,cm in length for head imaging. 
This field inhomogeneity works with an RF system with a bandwidth of $<$10\%.
The fields generated by the proposed PMA are validated using both analytic calculation and numerical simulation software. 
The PMA consists of a main array and a sub-gradient-array. The main array supplies a strong and relatively homogeneous magnetic field and it bears the main weight (120.96\,kg), whereas the sub-gradient-array supplies a monotonic field pattern and it is light (5.12\,kg). 
It has a magnetic field generation efficiency of 0.88\,mT/kg which is the highest among sparse PMAs that offer a monotonic field pattern.
The proposed PMA can be used to supply gradient fields in a single direction working with gradient coils in the other two directions, or be rotated to encode signals for imaging with an axial selection.
The encoding capability of the magnetic field is validated using numerical simulations where the field pattern of the proposed PMA even outperforms either a linear pattern or the patterns supplied by other PMA's.
The rotation of the field pattern can be achieved by rotating the much lighter sub-gradient-array, which significantly lowers the load of the mechanical rotation system, leading to a reduction in rotation error. 
For the optimization of the PMA, a physics-guided fast GA was used where physics-guided design with pre-design inspections and design synthesizer is introduced to narrow down the search landscape for GA optimizations. The GA is accelerated by a fast forward magnetic field calculation supported by the in-house-built code, name ``MagCraft''.
The axial direction of the magnetic field of the proposed magnet is the same as that of most of the superconducting magnets for MRI. This allows easy adoptions of the advancement of RF coils to the imaging system such as surface coils and a multi-channel system without compromising coil efficiency. 
In terms of implementation, the magnet is sparse, and is composed of permanent magnet blocks that are available off the shelf, which makes it ease to build. The force each magnet experiences in the design was calculated to confirm the feasibility of assembly. 
Moreover, the proposed PMA has 5-gauss range of 87\,$\times$\,87\,$\times$\,104\,cm$^3$,
thus it can be operated in a small space. 
With the characteristics of permanent magnets which are no power consumption and no need for a cooling system, and the fact that the magnet array can supply partial gradient fields to reduced the number of gradient coils, the proposed PMA can offer an MRI system with simplified hardware and a significant reduction in power consumption. It can be a good alternative for a portable MRI head imaging system. The design can be scaled and varied for imaging other parts of the human body.  

\section*{Acknowledgment}
Ting-Ou Liang would like to acknowledge the President Scholarship from Singapore University of Technology and Design for her Ph.D. study.

\bibliographystyle{IEEEtran}
\bibliography{sample}

\input{main_app}

\end{document}

%% file: fig_3D2D_view.tex
\begin{figure*}[t]
\centering
\begin{center}
\newcommand{\patchSize}{2.45cm}
\scriptsize
\setlength\tabcolsep{0.1cm}
\includegraphics[width=0.8\linewidth]{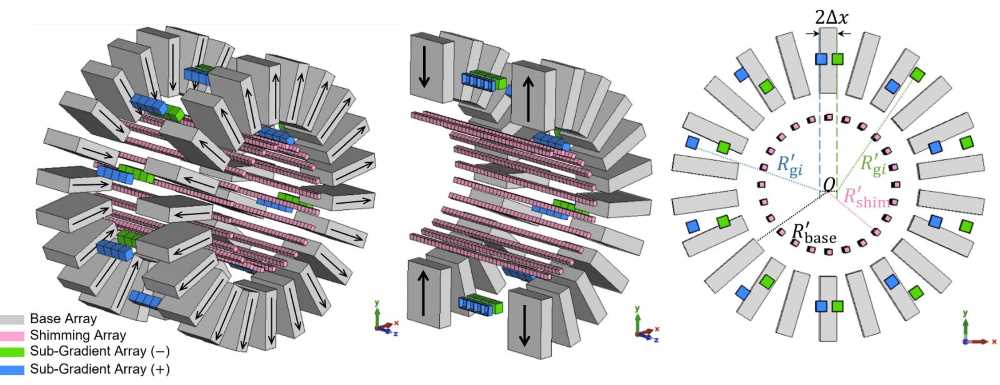}\\
(a)\hspace{0.23\linewidth}(b)\hspace{0.23\linewidth}(c)\\    	
\end{center}
\caption{The proposed array which is composed of the main array and a sub-gradient array (in blue and green), where the main array is a combination of the IO ring pair (also named base array, in grey), a shimming array (in red), and the iron bars connecting the IO ring pair (in dark grey), (a) 3D view, (b) the cutaway view that is cut through the $yz$-plane (blue), (c) the cross-sectional view on the $xy$-plane.}
\label{fig:array}
\end{figure*}

%% file: fig_magnet_block_current_model.tex
\begin{figure*}[t]
\centering
\begin{center}
\newcommand{\patchSize}{2.45cm}
\scriptsize
\setlength\tabcolsep{0.1cm}
\hspace{0.015\linewidth}
\includegraphics[width=0.25\linewidth]{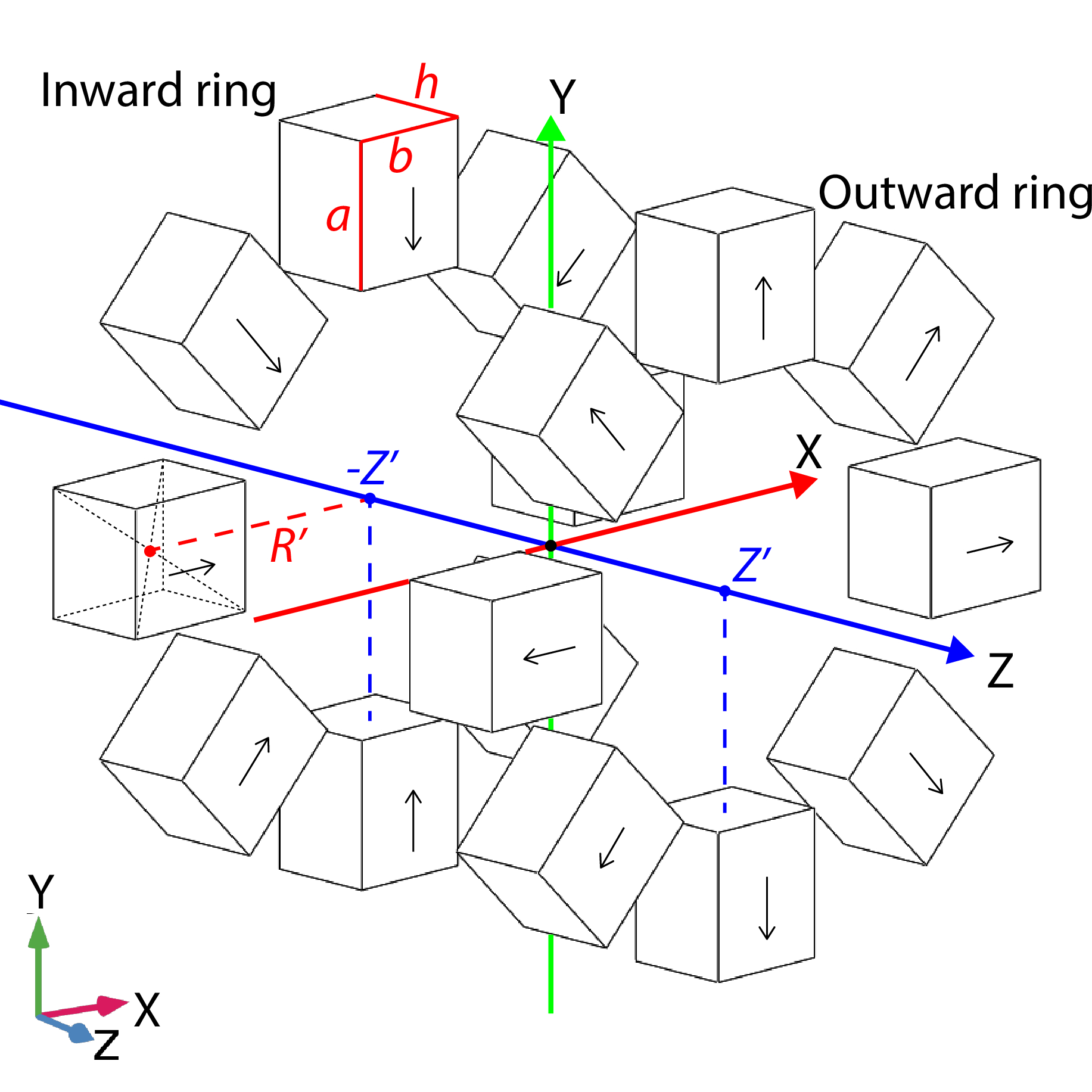}
\hspace{0.1\linewidth}
\includegraphics[width=0.25\linewidth]{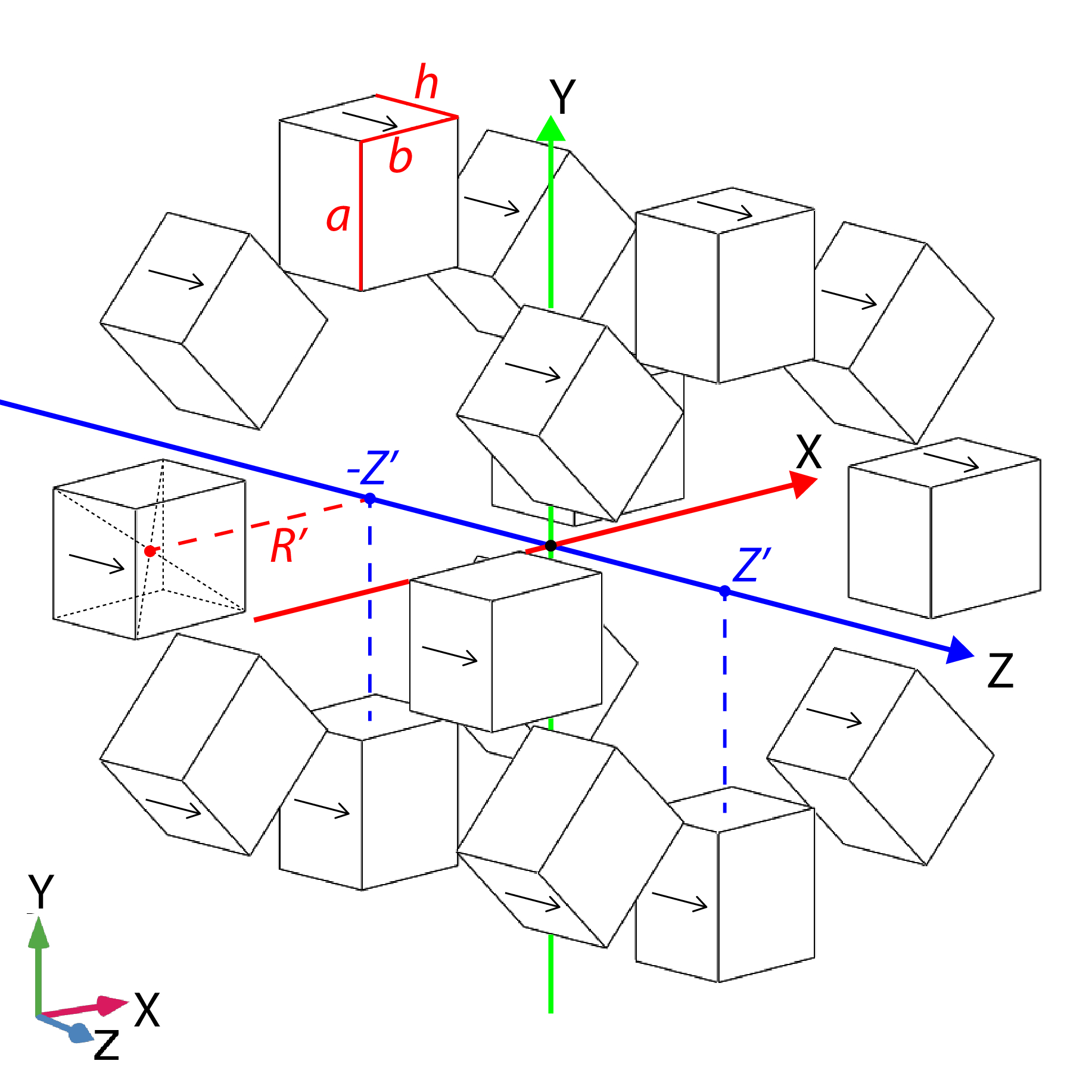}\\
(a)\hspace{0.35\linewidth}(b) \\   
\end{center}
\caption{(a) An IO type ring pair, (b) a parallel type ring pair.}
\label{fig:block_current_model}
\end{figure*}

%% file: fig_3D_overview.tex
\begin{figure*}[t]
\centering
\begin{center}
		\newcommand{\patchSize}{2.45cm}
		\scriptsize
		\setlength\tabcolsep{0.1cm}
    	\includegraphics[width=0.7\linewidth]{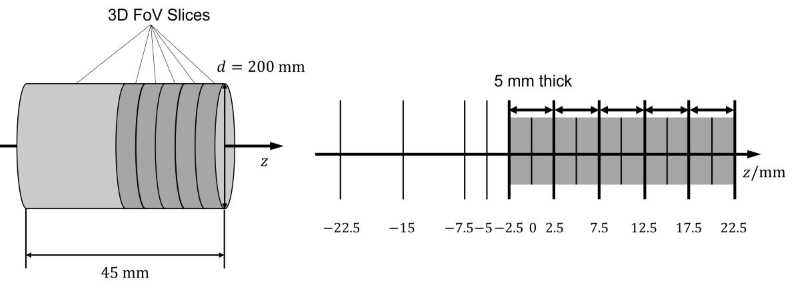}\\
        (a)\hspace{0.3\linewidth}(b)\hspace{0.12\linewidth}\phantom{(c)}\\
\end{center}
\caption{(a) The overview of a 3D FoV for imaging; (b) The $z$-slices selected for inspection in this paper. There are five cylinders with a thickness of 5 mm, which are indicated by the grey boxes in (b). The region of interest is of 45 mm thickness.}
\label{fig:3D_overview}
\end{figure*}

%% file: fig_flowchart.tex
\begin{figure*}[t]
\centering
\includegraphics[width=0.8\linewidth]{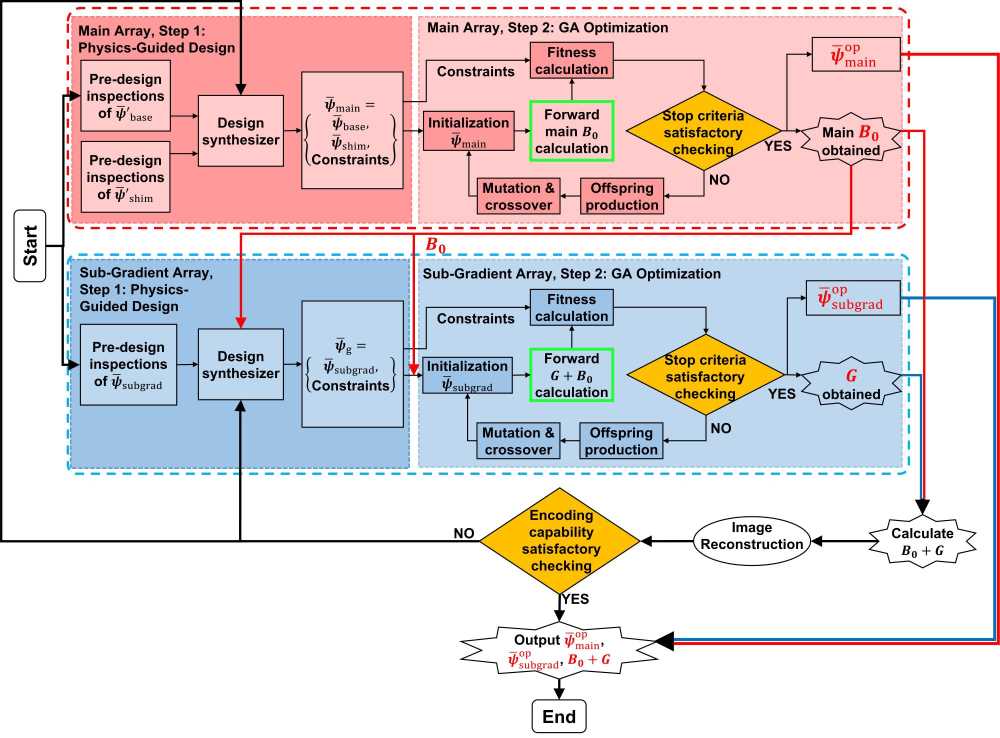}
\caption{The optimization flow of the proposed magnet array. The red and the blue blocks show the optimizations of the main array and the sub-gradient array, respectively. Each block includes two steps, the physics-guided design and the GA optimization. The current model was used for the forward calculation of all the steps. The optimized main field $B_0$ is input to both the physics-guided design and GA optimization of the sub-gradient array. In addition, there is a feedback loop from the image reconstruction to the physics-guided design of sub-gradient array to refine the constraint for optimization. }
\label{fig:flowchart}
\end{figure*}

%% file: fig_main_array.tex
\begin{figure*}[t]
\centering
\begin{center}
		\newcommand{\patchSize}{2.45cm}
		\scriptsize
		\setlength\tabcolsep{0.1cm}
    	\hspace{0.015\linewidth}
    	\includegraphics[width=0.8\linewidth]{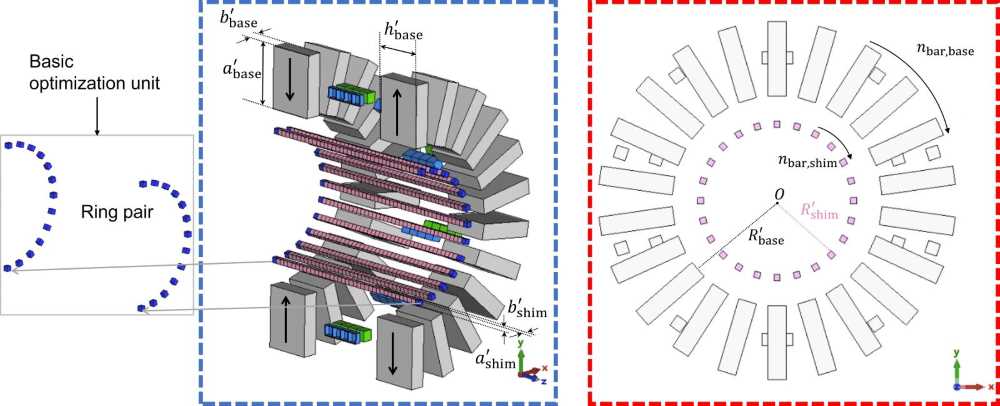}\\
        \hspace{0.1\linewidth}(a)\hspace{0.4\linewidth}(b)\hspace{0.5\linewidth}\\
		\includegraphics[width=0.8\linewidth]{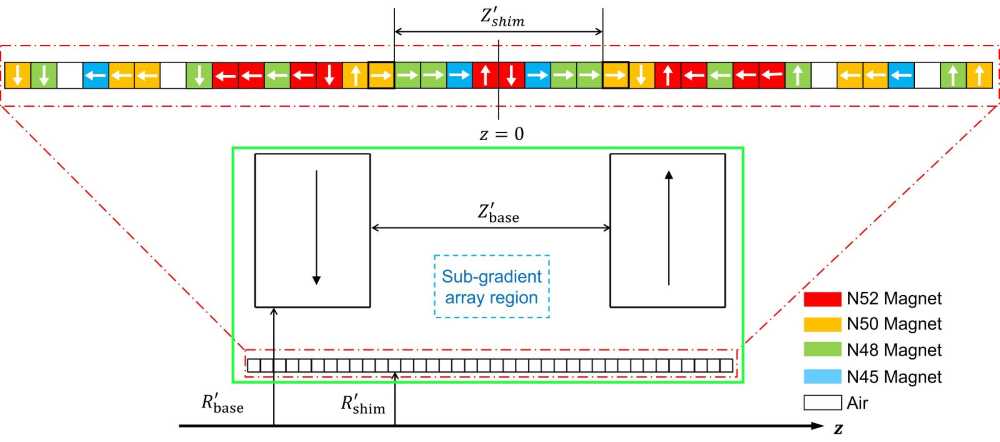}\\
		(c)
\end{center}
\caption{The main array, (a) cutaway view cut through the $yz$-plane, (b) 2D front view, (c) side view (above the z-axis) with the polarizations of magnets in the shimming array shown.}
\label{fig:main_aray}
\end{figure*}

%% file: fig_predesign.tex
\begin{figure*}[t]
\centering
\includegraphics[width=0.8\linewidth]{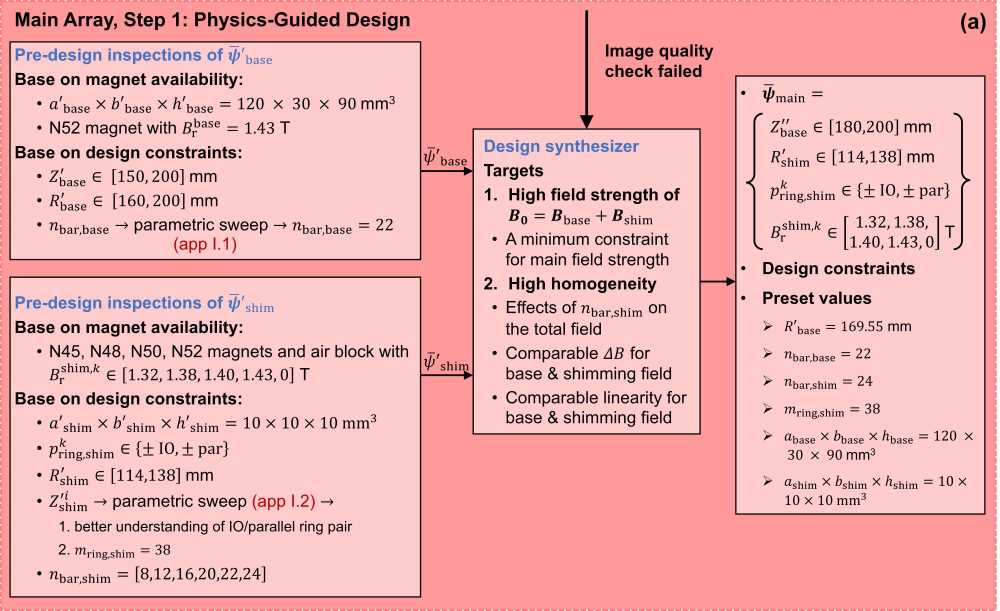}\\
\includegraphics[width=0.8\linewidth]{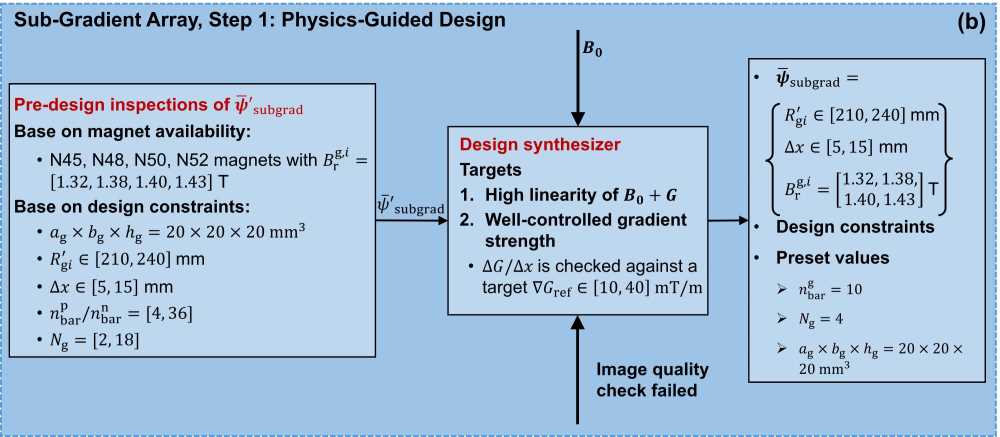}\\
\caption{The detail of the physics-guided design of (a) the main array and (b) the sub-gradient array. In the pre-design inspection blocks (left most), an initial range of all the parameters for each sub-array are determined separately by either the availability of magnets and the design constraint. In a design synthesizer (in the middle), the design parameters are further tailored toward the optimization targets. The right most blocks show the outputs, including the parameters with refined ranges, the design constraints, and the preset values.}
\label{fig:predesign}
\end{figure*}

%% file: fig_ring_field_current_model.tex
\begin{figure*}[t]
\centering
\begin{center}
		\newcommand{\patchSize}{2.45cm}
		\scriptsize
		\setlength\tabcolsep{0.1cm}
       \includegraphics[width=0.9\linewidth]{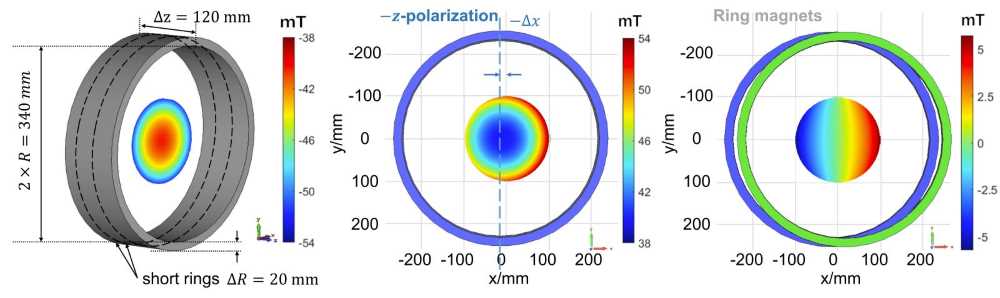}\\
        \hspace{0.05\linewidth}(a)\hspace{0.28\linewidth}(c)\hspace{0.28\linewidth}(e)\\
    	\includegraphics[width=0.9\linewidth]{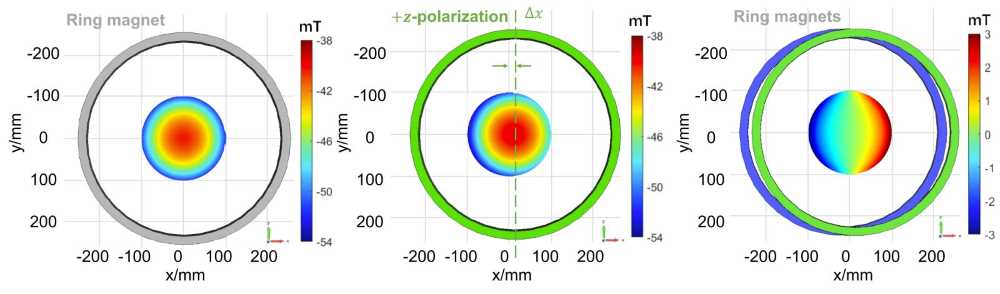}\\
        \hspace{0.05\linewidth}(b)\hspace{0.28\linewidth}(d)\hspace{0.28\linewidth}(f)\\
\end{center}
\caption{A ring magnet/magnets (a remanence of $B_\text{r}$ = \SI{1.4}{\tesla}, $R$\,=\,230\,mm, $\Delta R$\,=\,20\,mm, $\Delta Z$\,=\,200\,mm for (a)-(e) and $\Delta Z$\,=\,10\,mm for (f)) and the field distribution of $B_z$ in a circle centered at the origin with a diameter of 100\,mm, (a) 3D view (b) side view of a magnetization in the $-z$ direction, (c) -$z$-polarization ring with an offset of $\Delta x$\,=\,8\,mm along the $-$x-direction, (d) +$z$-polarization ring with the same amount of offset along the $+$x-direction, (e) the offset-double-ring which is a combination of the rings in (c) and (d), (f) the offset-double-ring when the rings are shortened ($\Delta Z$\,=\,10\,mm).}
\label{fig:ring_magnet}
\end{figure*}

%% file: fig_subgradient_array.tex
\begin{figure*}[t]
\centering
\begin{center}
		\newcommand{\patchSize}{2.45cm}
		\scriptsize
		\setlength\tabcolsep{0.1cm}
        \includegraphics[width=0.8\linewidth]{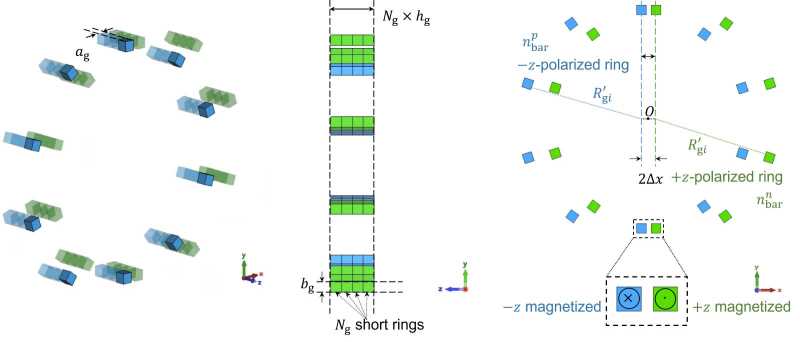}\\
        \hspace{0.05\linewidth}(a)\hspace{0.27\linewidth}(b)\hspace{0.28\linewidth}(c)\\
\end{center}
\caption{The sub-gradient array, (a) 3D cutaway view cutting through the $yz$-plane, (b) 2D side view, and (c) 2D front view.}
\label{fig:subgradient_array}
\end{figure*}

%% file: fig_field_3D.tex
\begin{figure*}[t]
\centering
\begin{center}
		\newcommand{\patchSize}{2.45cm}
		\scriptsize
		\setlength\tabcolsep{0.1cm}
        \includegraphics[width=0.9\linewidth]{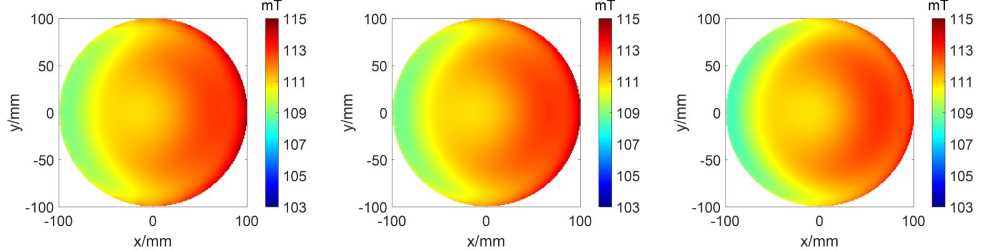}\hspace{0.5\linewidth}\\
        \hspace{0.05\linewidth}(a)\hspace{0.28\linewidth}(b)\hspace{0.28\linewidth}(c)\\
        \includegraphics[width=0.6\linewidth]{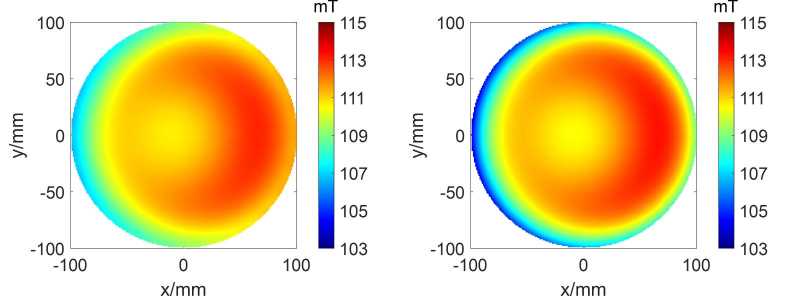}\hspace{0.3\linewidth}\\
        \hspace{0.05\linewidth}(d)\hspace{0.28\linewidth}(e)\hspace{0.28\linewidth}\\
\end{center}
\caption{The field distributions of $B_\text{z}$ of (a) $z=0$ mm, (b) $z=5$ mm, (c) $z=10$ mm, (d) $z=15$ mm, and (e) $z=20$ mm.}
\label{fig:field_3D}
\end{figure*}

%% file: tab_result_3D.tex
\begin{table*}[ht]
    \small
    \centering
    \begin{tabular}{|c|c|c|c|c|}
        \hline
        \makecell{\textbf{Slice} \\ \textbf{ index}} & \makecell{\textbf{$z_\text{center}$} \\ \textbf{/\si{\milli\meter}}} & \makecell{\textbf{mean$(B_0+G)$} \\ \textbf{/\si{\milli\tesla}}} & \makecell{\textbf{$\Delta (B_0+G)$} \\ \textbf{(5-mm-thick 3D)/\si{\milli\tesla}}} & \makecell{\textbf{$\Delta (B_0+G)$} \\ \textbf{(2D)/\si{\milli\tesla}}} \\
        \hline
        \textbf{1} & 0  & 111.40 & 5.94 & 5.46 \\ \hline
        \textbf{2} & 5  & 111.33 & 6.01 & 5.77 \\ \hline
        \textbf{3} & 10 & 111.16 & 6.10 & 5.99 \\ \hline
        \textbf{4} & 15 & 110.83 & 9.17 & 9.09 \\ \hline
        \textbf{5} & 20 & 110.79 & 10.57 & 10.57 \\ \hline
    \end{tabular}
    \caption{The charateristics of the magnetic fields of slices at $z=0$\,mm in Fig.\,\ref{fig:field_map_total}\,(a), and $z=5$, 10, 15, and 20\,mm in Fig.\,\ref{fig:field_3D}. Column\,4 presents $\Delta (B_0+G)$ of the 5\,mm-thick 3D slice and Column\,4 presents that of the 2D center slide.}
    \label{tab:result_3D}
\end{table*}

%% file: tab_subgrad_detail.tex
\begin{table*}[t]
\small
\begin{center}
    \caption{The optimized inner radii and remanence for each column of the sub-gradient array.}
    \begin{tabular}{|c|c|c|c|c|c|c|c|c|c|c|} 
    \hline
    \textbf{Index} & \footnotesize{\textbf{1}} & \footnotesize{\textbf{2}} & \footnotesize{\textbf{3}} & \footnotesize{\textbf{4}} & \footnotesize{\textbf{5}} & \footnotesize{\textbf{6}} &\footnotesize{\textbf{ 7}} & \footnotesize{\textbf{8}} & \footnotesize{\textbf{9}} & \footnotesize{\textbf{10}} \\ [0.5ex] 
    \hline
    $R_\text{g}$/mm & 221 & 240 & 240 & 239 & 236 & 214 & 211 & 211 & 211 & 213\\[0.5ex]
    \hline
    $B_\text{r}$/T & 1.32 & 1.32 & 1.38 & 1.40 & 1.38 & 1.38 & 1.40 & 1.40 & 1.40 & 1.38\\[0.5ex] 
    \hline
    \end{tabular}
    \label{tab:subgrad_detail}
\end{center}
\end{table*}

%% file: fig_field_map_total.tex
\begin{figure*}[t]
\centering
\begin{center}
		\newcommand{\patchSize}{2.45cm}
		\scriptsize
		\setlength\tabcolsep{0.1cm}
        \includegraphics[width=0.6\linewidth]{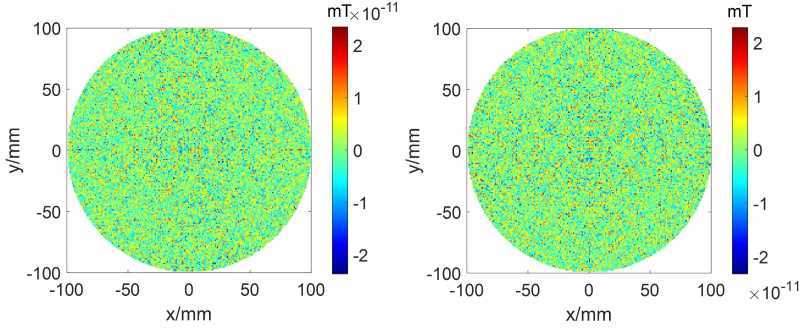}\\
        (a)\hspace{0.3\linewidth}(b)\\
\end{center}
\caption{The field distributions of (a) $B_x$ and (b) $B_y$ of the total magnetic field of the center slice at $z=0$\,mm of the proposed PMA.}
\label{fig:field_map_total}
\end{figure*}

%% file: fig_field_maps2.tex
\begin{figure*}[t]
\centering
\begin{center}
		\newcommand{\patchSize}{2.45cm}
		\scriptsize
		\setlength\tabcolsep{0.1cm}
        \includegraphics[width=0.9\linewidth]{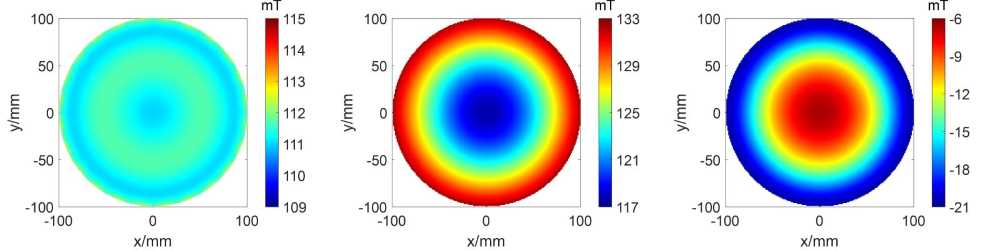}\hspace{0.5\linewidth}\\
        \hspace{0.05\linewidth}(a)\hspace{0.28\linewidth}(b)\hspace{0.28\linewidth}(c)\\
        \includegraphics[width=0.6\linewidth]{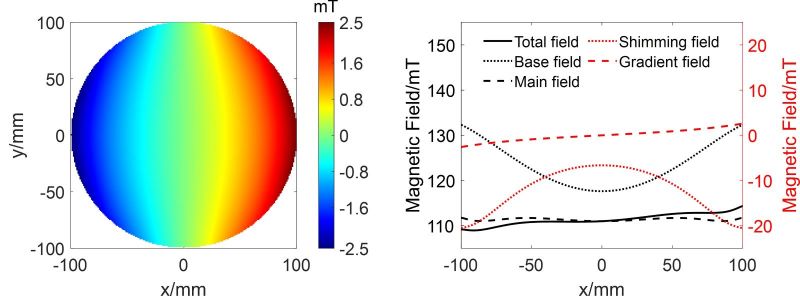}\hspace{0.3\linewidth}\\
        \hspace{0.05\linewidth}(d)\hspace{0.3\linewidth}(e)\hspace{0.3\linewidth}\\
\end{center}
\caption{The field distributions of $B_z$ of the center slide of (a) the main array (average field strength: 111.40\,mT, inhomogeneity: 1.27\,mT) (b) the base array (average field strength: 126.06\,mT, inhomogeneity: 14.70\,mT), (c)shimming array, (d) the sub-gradient array, and (e) the 1D field plots on $x$-axis for \ref{fig:field_map_total}\,(a) and \ref{fig:field_map2}\,(a)-(d).}
\label{fig:field_map2}
\end{figure*}

%% file: fig_simulated_images_3D.tex
\begin{figure*}[t]
\centering
\begin{center}
		\newcommand{\patchSize}{2.45cm}
		\scriptsize
		\setlength\tabcolsep{0.1cm}
        \includegraphics[width=0.8\linewidth]{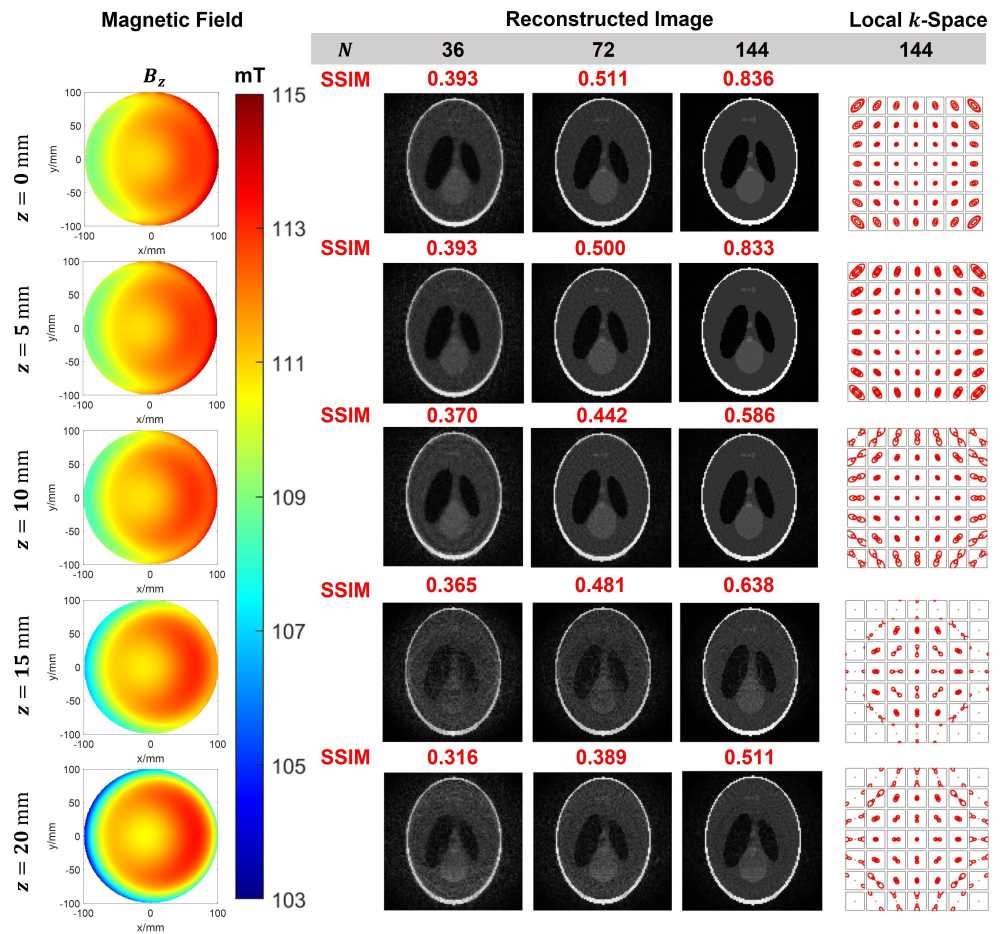}
\end{center}
\caption{The reconstructed images of the Shepp-Logon phantom (middle, column\,2-4) by using the magnetic field supplied by the proposed PMA on different slices: $z=0,5,10,15,\text{ and }20$ mm when the number of rotation angles is varied. The corresponding local k-spaces of the field patterns at the number of rotation angles 144 are shown in Column\,5. }
\label{fig:simulated_images_3D}
\end{figure*}

%% file: fig_simulated_images.tex
\begin{figure*}[t]
\centering
\begin{center}
		\newcommand{\patchSize}{2.45cm}
		\scriptsize
		\setlength\tabcolsep{0.1cm}
        \includegraphics[width=0.8\linewidth]{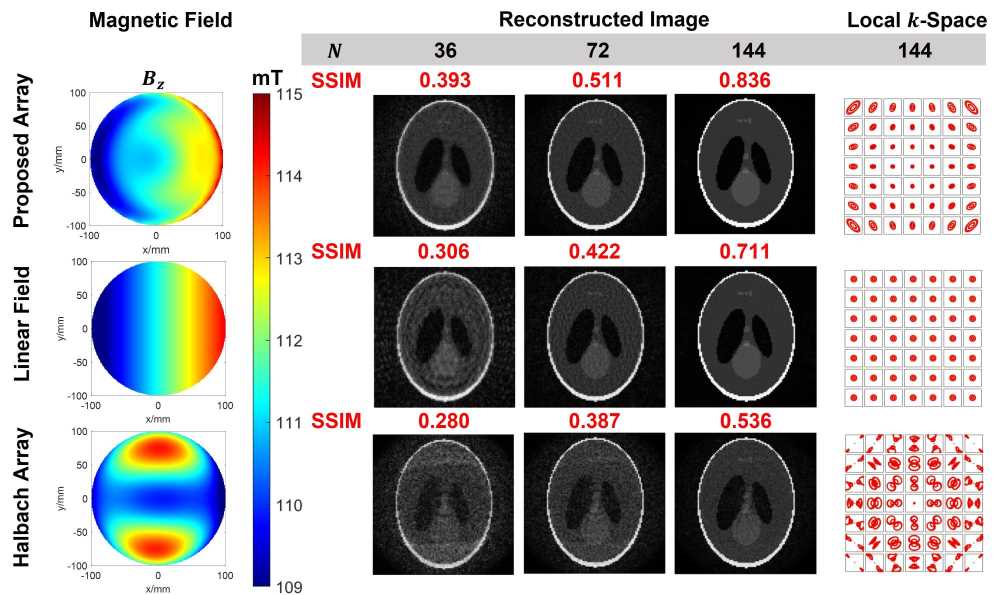}
\end{center}
\caption{ The reconstructed images of the Shepp-Logon phantom (middle, column\,2-4) by using the magnetic field supplied by the proposed PMA, a linear field, and a field by a short Halbach Array (shown in the first column) when the number of rotation angles is varied. The corresponding local $k$-spaces of the field patterns at the number of rotation angles 144 are shown in Column\,5. }
\label{fig:simulated_images}
\end{figure*}

%% file: fig_PSF.tex
\begin{figure*}[t]
    \centering
    \includegraphics[width=0.8\linewidth]{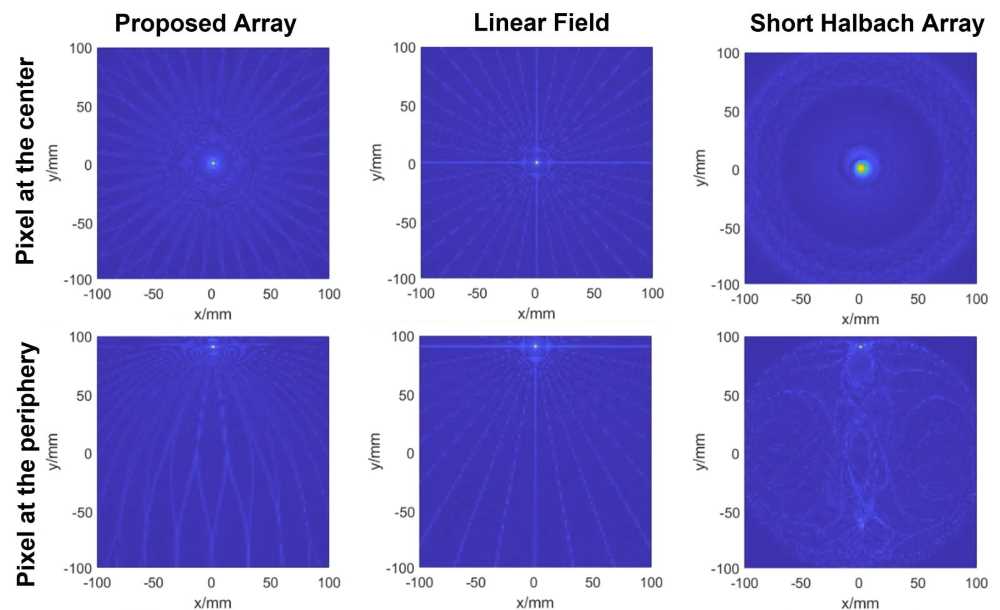}
    \caption{The images with the PSF's of the two pixels, one at the center (0,0)\,(mm) and the other at the periphery of the phantom (0,90)\,(mm) in an MRI system using the field pattern supplied by the proposed PMA, a linear pattern and a Halbach pattern shown in the first column in Fig.\,\ref{fig:simulated_images}. The coil sensitivity was set to be uniform. The number of rotation angles was set to $N$\,=\,36 for the six sub-figures. The three columns are for the pattern from the proposed PMA, the linear pattern, and the Halbach pattern, respectively. The first row is for the pixel at the center and the second row is for the pixel at the periphery of the phantom.} 
    \label{fig:fig_PSF}
\end{figure*}

%% file: fig_5_Gauss.tex
\begin{figure*}[t]
\centering
\begin{center}
		\newcommand{\patchSize}{2.45cm}
		\scriptsize
		\setlength\tabcolsep{0.1cm}
        \includegraphics[width=0.8\linewidth]{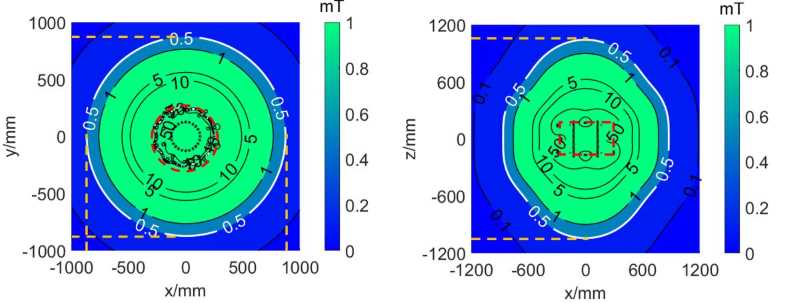}\\
        \hspace{0.01\linewidth}(a)\hspace{0.4\linewidth}(b)\\
\end{center}
\caption{The total field strength $|B|=\sqrt{B_x^2+B_y^2+B_z^2}$ on (a) $x$-$y$ plane and (b) $x$-$z$ plane. The 5-Gauss lines, i.e., 0.5 mT are highlighted with white color in both plots. In (a), the FoV is a square with both $x$ and $y$ ranging from $-1000$ to $1000$; in (b)(d), the FoV is a square with with $x$ and $z$ ranging from $-1200$ to $1200$.}
\label{fig:5Gauss}
\end{figure*}

%% file: main_app.tex
\appendices

\section*{Appendix \RomanNumeralCaps{1}: Parametric sweeps for pre-design inspections and design synthesizer of the main array}
\label{sec:append_main_array_design}

\input{fig_base_nbar}

\subsection*{\RomanNumeralCaps{1}.1: Parametric sweep of $n_\text{bar,base}$ and $Z'_\text{base}$}
\label{subsec:append_base_nbar}
The base array only have one IO ring pair as shown in Fig.\,\ref{fig:main_aray}\,(a) and (b). Around each ring, $n_\text{bar,base}$ is the number of magnet bars, as shown in Fig.\,\ref{fig:main_aray}\,(b). 
To inspect the effect of $n_\text{bar,base}$, the IO rings with different values of $n_\text{bar,base}$ were simulated using the proposed accelerated magnet simulator, SUTD-MagCraft (please refer to the confidential supplementary material for details). Fig.\,\ref{fig:base_nbar}\,(a) and (b) shows the mean and the range of $B_\text{base}$ versus $n_\text{bar,base}$, with $Z'_\text{base}$ set to 200\,mm and $a_\text{base}\times b_\text{base}\times h_\text{base}=120\times 30\times 90$\,mm$^3$. Both Fig.\,\ref{fig:base_nbar}\,(a) and (b) show that when $n_\text{bar,base}$ increases, both mean\,($B_\text{base}$) and $\triangle B_\text{base}$ increase linearly.

\input{fig_base_1D.tex}

Next, the effect of $Z'_\text{base}$ on the field pattern is inspected with a fixed $n_\text{bar,base}$. Fig.\,\ref{fig:base_1D} shows the calculated $B_\text{base}$ along the x-axis at $y$\,=\,0\,mm when $n_\text{bar,base}$ is fixed at 22 and $Z'_\text{base}$ is varied from 150\,mm to 200\,mm considering the possibility to accommodate the defined FoV in this paper. As the base array supplies a concentric field pattern, the 1D plots in Fig.\,\ref{fig:base_1D} are sufficient to provide indications on the pattern of gradient fields the magnet array can offer. As shown in Fig.\,\ref{fig:base_1D}, all the cases show a monotonical increase when moving from the center to the peripheral, i.e., from $x$\,=\,0 to $x$\,=\,100 or -100 mm. Linear regressions were performed on the curves at the positive side of Fig.\,\ref{fig:base_1D}, and the coefficient, $R^2$, was used as the indicator of linearity. The $R^2$ goes from 0.9484 to 0.9722 with $Z_\text{base}$ increases from 150 mm to 200 mm.


\subsection*{\RomanNumeralCaps{1}.2: Detailed study of the parameters of the shimming array}
\label{subsec:append_shim_param}

\input{Fig_single_ring_pair}

The shimming array consists of $m_\text{ring,\,shim}$ rings (i.e., $m_\text{ring,\,shim}$/2 ring pairs) and $n_\text{bar,\,shim}$ magnet bars around each ring as shown in Fig.\,\ref{fig:main_aray}\,(c) and (b), respectively. As introduced in the main content, the axial distance between the inner edges of the two rings of a ring pair is denoted as $Z'_\text{shim}$. The ring pair types include IO type (shown in Fig.\,\ref{fig:block_current_model}\,(a)) and parallel type (shown in Fig.\,\ref{fig:block_current_model}\,(b)). The field patterns of these two types of ring pairs at different $Z'_\text{shim}$s are shown in Fig.\,\ref{fig:single_ring_pair}. 
$Z'_\text{shim}$ has a range of [0, 360]\,mm when the maximum $Z'_\text{base}$ is set to be 200\,mm and thus the maximum of $m_\text{ring,\,shim}$ is 38 (i.e., 19 ring pairs) when 10\,mm magnet cubes are used. 
Fig.\,\ref{fig:single_ring_pair}\,(a) and (b) show the field patterns of the IO type ring pair when $Z'_\text{shim}$\,=\,0\,mm and 360\,mm, respectively, whereas Fig.\,\ref{fig:single_ring_pair}\,(c) and (d) show those of the parallel type ring pair when $Z'_\text{shim}$\,=\,0\,mm and 360\,mm, respectively. In Fig.\,\ref{fig:single_ring_pair}, it is seen that all cases show concentric field patterns. on the other hand, the two types of rings show different gradient patterns. Moreover, the pattern and strength is severely affected by the distance between the two rings. When the rings are near to each other, comparing Fig.\,\ref{fig:single_ring_pair}\,(a) and (c), it is seen that the parallel type of ring pair has much higher field compared to the IO type, whereas when the rings are far apart, 
as shown in Fig.\,\ref{fig:single_ring_pair}\,(b) and (d), both cases show a significant decrease in field strength, meanwhile, comparing the two, it can be observed that the IO-ring pair has much higher field strength. 
The variations due to the type of the ring pair and $Z'_\text{shim}$ offers flexibility to the optimization of the shimming array when it is design. Moreover, a higher $m_\text{ring,\,shim}$ leads to a higher degree of freedom in the design. One thing to take note is that when $m_\text{ring,\,shim}$ is high that the shimming array is longer than the base array, it may affect the access to the bore of the magnet. One the other hand, it may not be necessary as the contribution to the FoV become small when the rings are further apart. 

\input{fig_shim_sweep.tex}

The effect on the field pattern of the number of magnet blocks around a ring in the shimming sub-array, $n_\text{bar,shim}$, is investigated through GA optimizations for a combined field of the base and the shimming array when $n_\text{bar,shim}$ was set to be 8,\,12,\,16,\,20,\,22,\,24. The number of rings ($m_\text{ring,shim}$) was fixed at 38 where the end-to-end length of the shimming array is the same as that of the base array when $Z'_{base}$ is set to the maximum, 200\,mm to have a high degree of freedom for optimization. For the base array, based on the previous investigations, $n_\text{bar,\,base}$ was set at 22 and $Z'_\text{base}\in$\,[150,\,200]\,mm.
The resulting main field patterns are shown in Fig.\,\ref{fig:shim_sweep} with the $\Delta B_\text{1D}$ over $x$-axis and the average field of each map labeled on top. The $\Delta B_\text{1D}$ is used instead of the overall $\Delta B$, because there can be oscillations at the periphery of the field patterns, such as in Fig.\,\ref{fig:shim_sweep}\,(a). Since the oscillations are confined in the peripheral region, while the imaging process mainly use the central region, $\Delta B_\text{1D}$ can better reflect the property of the encoding field. Fig.\,\ref{fig:shim_sweep}\,(a), (b), (c), and (d) have $n_\text{bar,shim}=8,12,16,20$, respectively, and they are plotted with the same scale and colormap. Fig.\,\ref{fig:shim_sweep}\,(e) and (f) are the cases where $n_\text{bar,shim}=22$ and $24$, which share another scale and colormap. Both colorbars have a range of 10 mT for better comparison. It can be observed that when $n_\text{bar,shim}$ is small, such as 8 or 12, the field pattern is bumpy in the peripheral region. Moreover, all the cases have a inhomogeneity over $x$-axis higher than 1 mT except for the $n_\text{bar,shim}=24$ case, where $\Delta B_\text{z,main}=0.98$\,mT (inhomogeneity of 0.9\%). 
From the data obtained in Fig.\,\ref{fig:shim_sweep}, it shows that a higher $n_\text{bar,shim}$ leads to a higher field homogeneity after an optimization. However, a high $n_\text{bar,shim}$ increases the space occupied and increases the weight. By balancing the field homogeneity and the size and weight of the shimming array, $n_\text{bar,shim}$ was set to 24.

\subsection*{\RomanNumeralCaps{1}.3: The matching of $\Delta B_\text{base}$ and $\Delta B_\text{shim}$}
\label{subsec:append_base_shim_match}

\input{fig_shimming_max_delta.tex}
Based on the results presented in the previous section, both the base and the shimming array supply concentric field patterns. When they are combined, the homogeneity of the combined field is decided by the matching of their $\Delta B_\text{base}$ and $\Delta B_\text{shim}$ as well as the matching of their linearity. The range of $\Delta B_\text{base}$ is from 13\,mT to 31\,mT when $Z'_\text{base}$ is varied from 150\,mm to 200\,mm as shown in Fig.\,\ref{fig:base_1D}\,(b) while for the same range of $Z'_\text{base}$, all the cases show a $R^2$ of higher than 0.945 for linearity as shown in Fig.\,\ref{fig:base_1D}\,(c). 
In this section, investigation is conducted to explore the range of $\Delta B_\text{shim}$ through a GA optimization as follows.

%

Two GA optimizations were run to explore a maximum $\Delta B_\text{shim}$, one without constrain and the other with constrains of $R^2>=0.90$ and field direction pointing to the $-z$ direction. The threshold of $R^2$ was chosen to match that of the base array. Although the base array shows a R$^2$ of higher than 0.945, it is set to be slightly lower to secure a large enough solution space.
In both optimizations, $m_\text{ring,shim}=38$ where the end-to-end length of the shimming array is the same as that of the base array when $Z'_{base}$ is set to the maximum, 200\,mm to have a high degree of freedom for optimization, and $n_\text{bar,shim}=24$ to have a high homogeneity of $B_\text{shim}$ based on the previous investigations. 
Fig.\,\ref{fig:shim_max_delta}\,(a) and (b) show the 2D field map and 1D field on $x$-axis of the unconstrained case, while Fig.\,\ref{fig:shim_max_delta}\,(c) and (d) show the results for the constrained one. 
As shown in Fig.\,\ref{fig:shim_max_delta}, the unconstrained case has $\Delta B_\text{shim}=98.76 \text{mT}$, while the constrained case has $\Delta B_\text{shim}=20.09 \text{mT}$. To match the linearity, $\Delta B_\text{shim}<20.09 \text{mT}$. Therefore, the linearity matching is translated to the constrains of $\Delta B<20 \text{mT}$. To match this range of $\Delta B$, based on the investigation of $\Delta B_\text{base}$ in Fig.\,\ref{fig:base_1D}\,(b), the lower bound of $Z'_\text{base}$ can accordingly be changed to 180\,mm. 


\input{fig_Taylor_coefficients.tex}
\section*{Appendix \RomanNumeralCaps{2}: the magnetic field of a magnet ring}
\label{sec:append_magnetRing_field}
For the $z$-component of the magnetic field supplied by a magnet ring as shown in Fig.\,\ref{fig:ring_magnet}, it can be expressed as $B_\text{z} (x,y)=\sum^N_{n=0}\;C_n\;(x^{2n}+y^{2n})$, where $C_n$ is the coefficient of the $n^\text{th}$ order term. Alternatively, in a cylindrical coordinate system, it can be expressed using the equation below,
\begin{equation}
\label{eq:ring_field}
B_\text{z} (\textbf{r}) = \frac{B_r}{4\pi}\int\displaylimits_{\Delta z} \int_{0}^{2\pi} \left(\sum^2_{i = 1}(-1)^i\frac{R_i^2-R_ir\cos\theta'}{(r^2+R_i^2-2rR_i\cos(\theta')+(z-z')^2)^{\frac{3}{2}}}\right) \mathrm {d}\theta'\mathrm {d}z'
\end{equation}
where $R_1 = R + \Delta R, R_2 = R$ are the outer and the inner radius of the ring, and $R$ and $\Delta R$ are labelled in Fig.\,\ref{fig:ring_magnet}. Cylindrical coordinate was used.
Taylor series was used to expand Eq.\,(\ref{eq:ring_field}) at $r = 0$. As the field pattern is concentric, $B_\text{z} (\textbf{r})$ could be reduced to a two-variable function $B_\text{z} (r, z)$ and Eq.\,(\ref{eq:ring_field}) is re-written as,
\begin{equation}\label{eq:taylor_series}
B_\text{z}(r, z)=\sum_{k=0}^\infty \frac{r^k}{k!}\frac{\partial^k}{\partial r^k}B_\text{z}(r, z)|_{r=0}
\end{equation}
where $k$ is an even number.
According to Leibniz integral rule, the coefficients in Eq.\,(\ref{eq:taylor_series}) is re-expressed as follows,
\input{eq_leibniz.tex}
where $b(r, z, \theta', z')$ is the integrand in Eq.\,(\ref{eq:ring_field}). Coefficients $C_1$ and $C_2$ correspond to the value of Eq. (\ref{eq:taylor_series}) when $k=2$ and $k=4$, respectively. In order to examine the condition when the higher order terms can be neglected, the ratio of $C_2/C_1$ is plotted with respect to $\Delta Z$ in Fig.\,\ref{fig:taylor_coefficients}. As can be seen in Fig.\,\ref{fig:taylor_coefficients}, when $\Delta Z$ is in the range of (235.9, 236.2)\,mm, $C_2/C_1$\,$<$\,0.01, and the fourth-order term is negligible. 


\section*{Appendix \RomanNumeralCaps{3}: Validation of the force calculation of magnet blocks}
\label{sec:append_force_cal_validation}

To validate the ``MagCraft'' for magnetic force calculation between permanent magnets, a simple example consisting of two magnet blocks is used to run CST simulation. The two magnet blocks are identical cubes with a side length of 10 mm. One is centered at the origin, and other is centered at $(x,y,z)=(0,30,0)$ (mm). They are both polarized in $+y$-direction with a remanence of $B_\text{r}=1.43 \si{\tesla}$. The force experienced by the two magnets are shown in Table\,\ref{tab:force_validate}. The resolution used for MagCraft calculation is 0.05 mm. Since the $x$- and $z$-components of the resulting magnetic forces are close to zero, only the $y$-component needs to be compared. If the CST simulation result is used as the reference, then the error of the MagCraft calculation should be 1.4\% and 1.2\%, respectively. 
When the magnitude of the force becomes stronger, the relative error will be further reduced. Since MagCraft is used to check the magnetic forces within the whole PMA, which can go up to several hundred Newton, an accuracy of about 1\% at a magnitude of 1 N can be considered as good.

\input{tab_force_validate.tex}

\section*{Appendix \RomanNumeralCaps{4}: GA Optimization for magnet insertion}
\label{sec:append_force_op}

In this section, the optimization of the magnet insertion sequence for the first base ring is presented in detail. This ring has 22 magnet blocks, therefore GA needs to select a 22-permutation that can effectively suppress the force experienced by each magnet during insertion. For each individual, the forces experienced by all magnets when they are being inserted to their designed positions are calculated. The maximum force is used as the fitness function, which aims for avoiding insertion of magnets that requires too large force. The population size is set to 50, and the stop criterion is the saturation of the best fitness value with a tolerance of 1\,N for 20 generations. Under this setting, GA stops at iteration 51. At the beginning, the average of the fitness value, that is, the average of the maximum forces during insertion over 50 random sequences, is 440\,N; however, at the end of the optimization, the resulting sequence gives the maximum force during insertion as 318\,N. This result shows an improvement of more than 25\% from a random sequence of magnet insertion, which comes out to be very effective.

\section*{Appendix \RomanNumeralCaps{5}: Experimental examinations of the temperature drift of N52 NdFeB magnet block}
\label{sec:append_temperature_drift}

\input{fig_temp_drift.tex}
\input{tab_temperature_drift.tex}

The temperature drift of the permanent magnet was tested on a 10\,mm N52 NdFeB cube magnet. Fig.\,\ref{fig:temp_drift} shows experimental setup. As shown in Fig.\,\ref{fig:temp_drift}, a Hall probe
(LakeShore 460 Gaussmeter) positioned at a PI (Physik Instrumente) xyz-moving platform (VT-80 linear
stage) was used to measure the $B_x$, $B_y$ and $B_z$ components of the magnetic field when the sensor tip was 10\,mm above the magnet cube, an actual distance of 12\,mm between the sensor inside the sensor tip and the magnet surface. 
On this plane, the measurement was conducted within a 2\,\si{\centi\meter}\,$\times$\,2\,\si{\centi\meter} region with a step size of 1\,\si{\centi\meter}. Five measurements were taken under a room temperature of \SI{22.9}{\celsius} and \SI{25.7}{\celsius}, and the average values of the measured magnetic field components were recorded in Table\,\ref{tab:temp_drift}. The temperature was measured using a Keysight U1181A immersion temperature probe. The ratio of the measured field strength between the two cases shows that the temperature drift of the magnetic field 12\,mm away from the magnet block is below 1\%.

%% file: fig_base_nbar.tex
\begin{figure*}[t]
\centering
\begin{center}
		\newcommand{\patchSize}{2.45cm}
		\scriptsize
		\setlength\tabcolsep{0.1cm}
    	\includegraphics[width=0.8\linewidth]{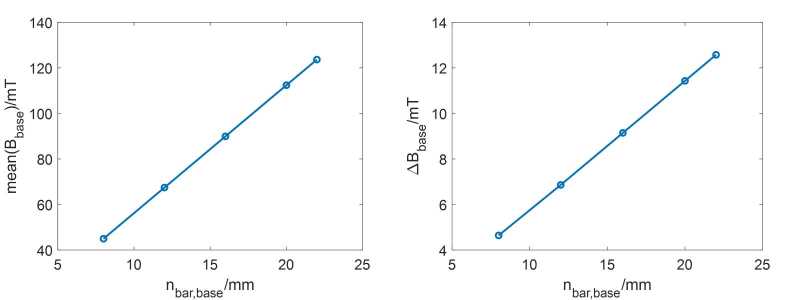}\\
    	\hspace{0.05\linewidth}(a)\hspace{0.4\linewidth}(b)
\end{center}
\caption{(a) The mean magnetic field strength and (b) the range of magnetic field strength within FoV with respect to the number of magnet bar pair, $n_\text{bar,base}$, for the base array. The remanence and inner radius of all magnet bars are fixed as $B_\text{r}=1.43$ and $R_\text{base}=169.55 \text{mm}$.This inner radius is the minimum value with $n_\text{bar,base}=22$. Also, the distance between the base IO-ring pair is set as $Z_\text{base}=200 \text{mm}$, which is the upper bound for this variable.} 
\label{fig:base_nbar}
\end{figure*}

%% file: fig_base_1D.tex
\begin{figure*}
\centering
\begin{center}
		\newcommand{\patchSize}{2.45cm}
		\scriptsize
		\setlength\tabcolsep{0.1cm}
    	\includegraphics[width=0.9\linewidth]{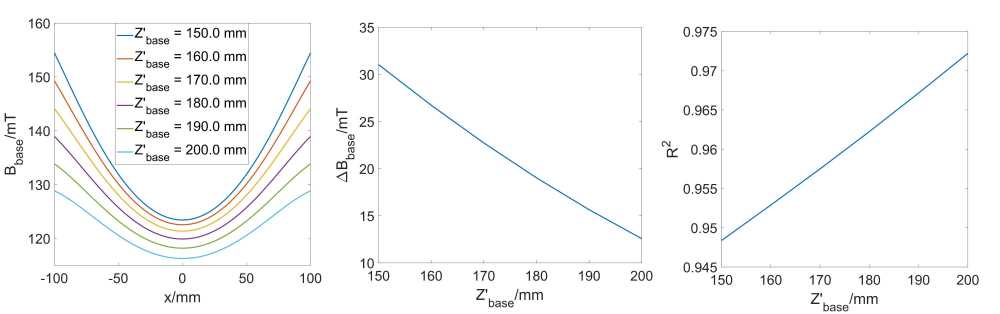}\\
    	\hspace{0.05\linewidth}(a)\hspace{0.3\linewidth}(b)\hspace{0.3\linewidth}(c)\hspace{0.2\linewidth}
\end{center}
\caption{The magnetic field on $x$-axis that is produced by the base array with $Z_\text{base}$ varying from 150 to 200 mm. The inner radius is fixed as 169.55 mm, and all magnets have the remanence of 1.43 T.} 
\label{fig:base_1D}
\end{figure*}

%% file: Fig_single_ring_pair.tex
\begin{figure*}[t]
\centering
\begin{center}
		\newcommand{\patchSize}{2.45cm}
		\scriptsize
		\setlength\tabcolsep{0.1cm}
    	\includegraphics[width=0.6\linewidth]{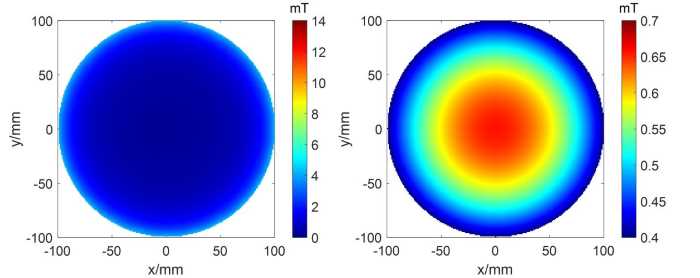}\\
    	\hspace{0.01\linewidth}(a)\hspace{0.25\linewidth}(b)\\
		\includegraphics[width=0.6\linewidth]{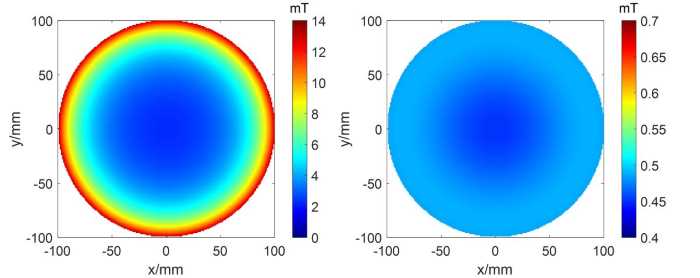}\\
		\hspace{0.01\linewidth}(c)\hspace{0.25\linewidth}(d)\\
\end{center}
\caption{The magnetic field of the IO type ring pair with (a) $Z'_\text{shim}=0$ mm, (b) $Z'_\text{shim}=360$ mm, the parallel type ring pair with  (c) $Z'_\text{shim}=0$ mm, (d) $Z'_\text{shim}=360$ mm. }
\label{fig:single_ring_pair}
\end{figure*}

%% file: fig_shim_sweep.tex
\begin{figure*}[t]
\centering
\begin{center}
		\newcommand{\patchSize}{2.45cm}
		\scriptsize
		\setlength\tabcolsep{0.1cm}
        \includegraphics[width=0.9\linewidth]{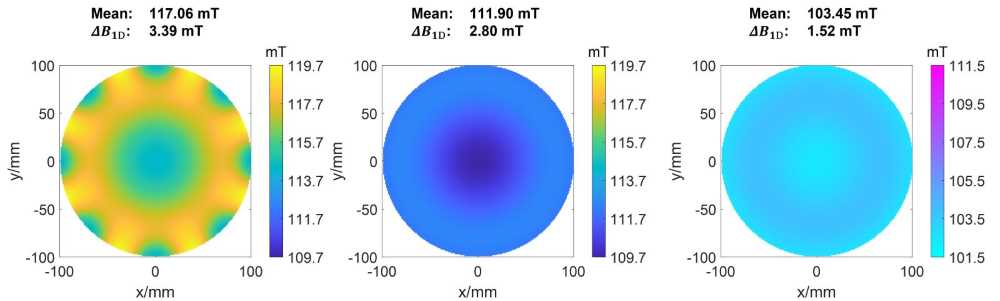}\\
        \hspace{0.01\linewidth}(a)\hspace{0.28\linewidth}(c)\hspace{0.28\linewidth}(e)\\
        \includegraphics[width=0.9\linewidth]{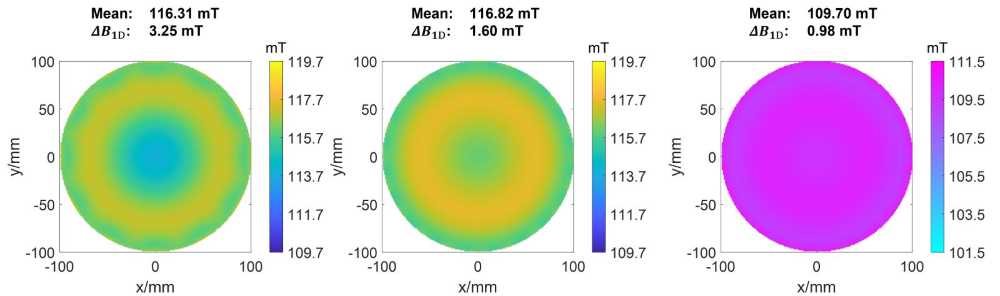}\\
        \hspace{0.01\linewidth}(b)\hspace{0.28\linewidth}(d)\hspace{0.28\linewidth}(f)\\
\end{center}
\caption{The optimized main magnetic field $B_\text{z}$ with (a) $n_\text{bar,shim}=8$, (b) $n_\text{bar,shim}=12$, (c) $n_\text{bar,shim}=16$, (d) $n_\text{bar,shim}=20$, (e) $n_\text{bar,shim}=22$, and (f) $n_\text{bar,shim}=24$. (a)-(d) are plotted in the same scale and colormap, while (e) and (f) are plotted in another scale and colormap. Both colorbars have a range of 10 mT.}
\label{fig:shim_sweep}
\end{figure*}

%% file: fig_shimming_max_delta.tex
\begin{figure*}[t]
\centering
\begin{center}
		\newcommand{\patchSize}{2.45cm}
		\scriptsize
		\setlength\tabcolsep{0.1cm}
    	\includegraphics[width=0.6\linewidth]{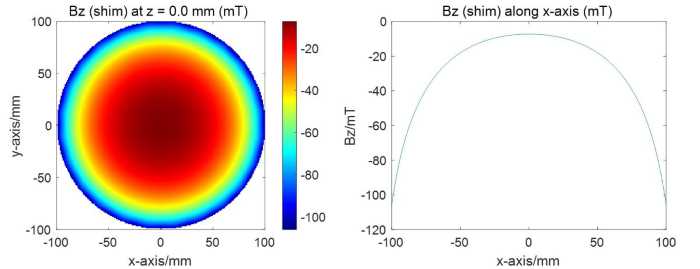}\\
    	\hspace{0.01\linewidth}(a)\hspace{0.25\linewidth}(b)\hspace{0.5\linewidth}\\
    	\includegraphics[width=0.6\linewidth]{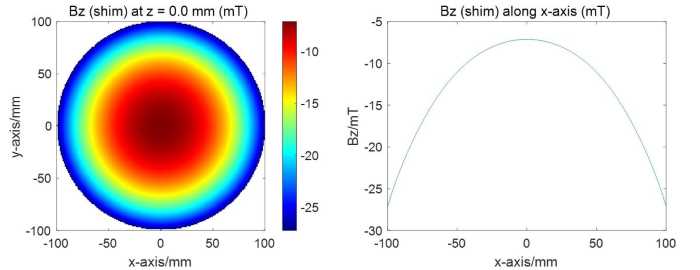}\\
    	\hspace{0.01\linewidth}(c)\hspace{0.25\linewidth}(d)\hspace{0.5\linewidth}
\end{center}
\caption{(a)(c) The 2D shimming field map, (b)(d) the 1D shimming field plots along the $x$-axis after the optimization that maximizes $\Delta B_\text{shim}$ where (a)(b) are the fields of the unconstrained case, while (c)(d) are those of the case that has the constraint that the coefficient of determination $R^2>=0.90$. The optimizations are run with $m_\text{ring,shim}=38$ and $n_\text{bar,shim}=24$.} 
\label{fig:shim_max_delta}
\end{figure*}

%% file: fig_Taylor_coefficients.tex
\begin{figure}[t]
\centering
\begin{center}
    \includegraphics[width=0.8\linewidth]{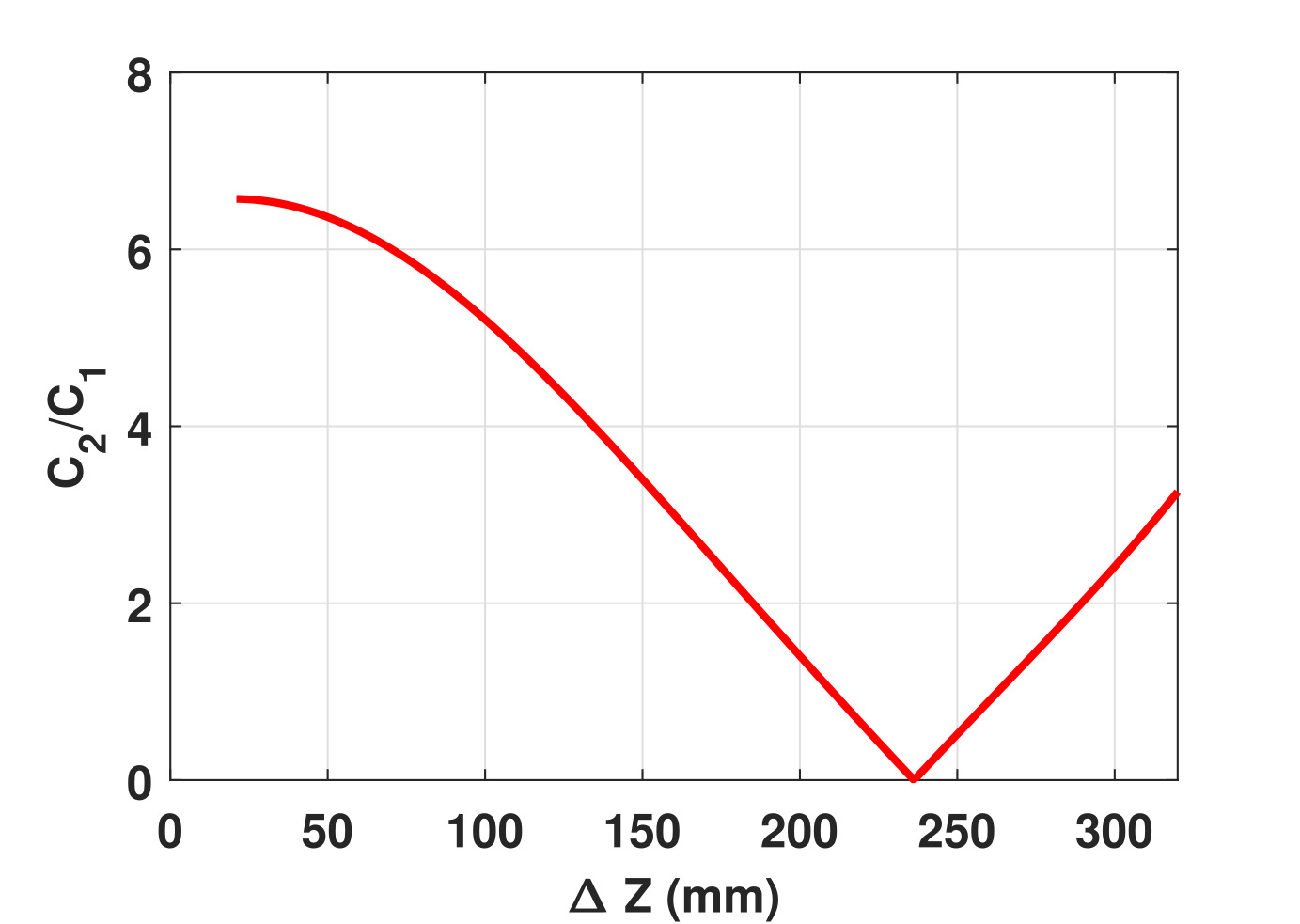}
\end{center}
\caption{The ratio of the fourth order coefficient over the second order coefficient, $C_2 / C_1$.}
\label{fig:taylor_coefficients}
\end{figure}

%% file: eq_leibniz.tex
\begin{equation}\label{eq:leibniz}
\begin{split}
    \frac{\partial^k B_\text{z} (r, z)}{\partial r^k} &= \frac{B_r}{4\pi}\frac{\partial^k}{\partial r^k} \int\displaylimits_{\Delta z'} \int_{0}^{2\pi} b(r, z, \theta', z') \mathrm {d}\theta '\mathrm {d}z'\\
    &= \frac{B_r}{4\pi}\int\displaylimits_{\Delta z'} \int_{0}^{2\pi}\frac{\partial^k}{\partial r^k} b(r, z, \theta', z') \mathrm {d}\theta '\mathrm {d}z'\\
\end{split}
\end{equation}

%% file: tab_force_validate.tex
\begin{table*}[ht]
    \small
    \centering
    \begin{tabular}{|c|c|c|c|c|}
        \hline
        \textbf{Method} & \textbf{Magnet index} & $F_\text{x}$/N & $F_\text{y}$/N & $F_\text{z}$/N\\
        \hline
        \multirow{2}{8em}{\textbf{CST simulation}}
        & \textbf{1} & -2.3924e-4 & 0.94307  & 4.9674e-4 \\
        & \textbf{2} & 2.3858e-4 & -0.94486  & 8.7691e-6 \\
        \hline
        \multirow{2}{8em}{\textbf{MagCraft calculation}}
        & \textbf{1} & -1.9062e-18 & 0.95640  & -2.6588e-19 \\
        & \textbf{2} & 1.3906e-17 & -0.95640 & -2.6588e-19 \\
        \hline
    \end{tabular}
    \caption{The forces between two identical magnet blocks computed by CST simulation and MagCraft calculation. The magnet blocks are cubes with a side length of 10 \si{\milli\meter} and polarization in $+y$-direction. The center locations of magnet 1 and 2 are $(x,y,z)=(0,0,0)$ (mm) and $(0,30,0)$ (mm), respectively. The remanence of the two magnets are $B_\text{r}=1.43$ \si{\tesla}.}
    \label{tab:force_validate}
\end{table*}

%% file: fig_temp_drift.tex
\begin{figure}[h]
\centering
    \begin{center}
        \includegraphics[width=0.8\linewidth]{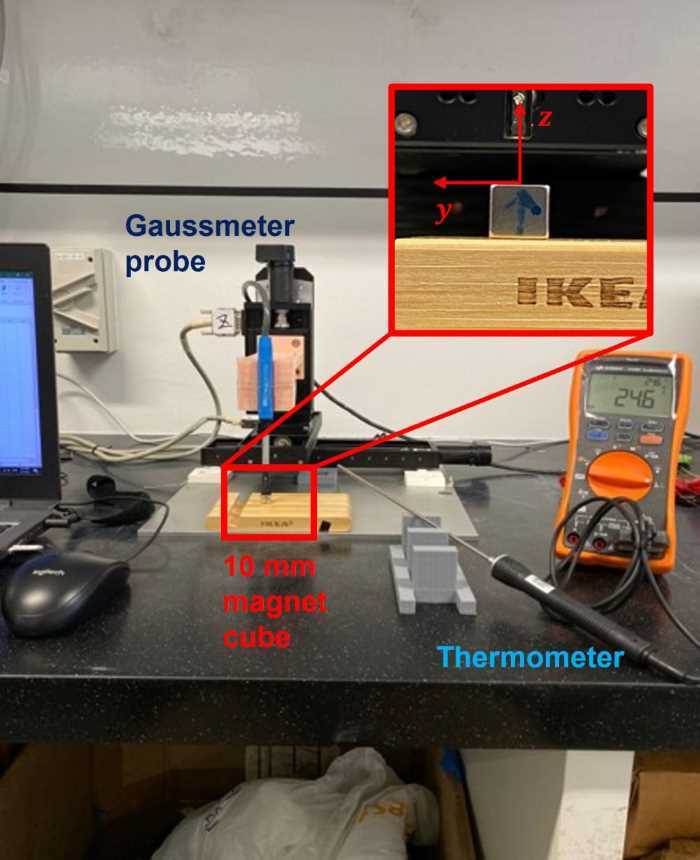}
    \end{center}
\caption{The hardware configurations for the magnetic field measurement to test the temperature drift of an N52 NdFeB magnet cube. The permanent magnet cube has a side length of 10 mm. The blue arrow in the close-up box indicates the polarization of the magnet.}
\label{fig:temp_drift}
\end{figure}

%% file: tab_temperature_drift.tex
\begin{table*}[h]
    \centering
    \footnotesize
    \begin{tabular}{|c c c|c c c|c c c|}
    \hline
         \multicolumn{3}{|c|}{\textbf{Case 1: \SI{22.9}{\celsius}}} & \multicolumn{3}{|c|}{\textbf{Case 2: \SI{25.7}{\celsius}}} & \multicolumn{3}{|c|}{\textbf{Ratio: Case 1/Case 2}}\\
         \hline
         \textbf{$B_\text{x}$/\si{\milli\tesla}} & \textbf{$B_\text{y}$/\si{\milli\tesla}} & \textbf{$B_\text{z}$/\si{\milli\tesla}} & \textbf{$B_\text{x}$/\si{\milli\tesla}} & \textbf{$B_\text{y}$/\si{\milli\tesla}} & \textbf{$B_\text{z}$/\si{\milli\tesla}} &
         \textbf{$R_\text{x}$} & \textbf{$R_\text{y}$} & \textbf{$R_\text{z}$}\\
         \hline
         9.615   & -11.931 & 8.439  & 9.627   & -11.928 & 8.381  & 0.999 & 1.000 & 1.007 \\ \hline
        -7.079  & -19.684 & 20.279 & -7.089  & -19.720 & 20.225 & 0.999 & 0.998 & 1.003 \\ \hline
        -12.118 & -10.454 & 7.558  & -12.106 & -10.453 & 7.494  & 1.001 & 1.000 & 1.008 \\ \hline
        21.450  & -5.801  & 19.998 & 21.564  & -5.848  & 19.987 & 0.995 & 0.992 & 1.001 \\ \hline
        -11.716 & -8.154  & 54.630 & -11.786 & -8.282  & 54.879 & 0.994 & 0.985 & 0.995 \\ \hline
        -19.888 & -1.775  & 18.346 & -19.932 & -1.822  & 18.336 & 0.998 & 0.974 & 1.001 \\ \hline
        12.455  & 10.194  & 7.790  & 12.500  & 10.219  & 7.774  & 0.996 & 0.998 & 1.002 \\ \hline
        -3.658  & 23.377  & 18.397 & -3.688  & 23.472  & 18.428 & 0.992 & 0.996 & 0.998 \\ \hline
        -10.732 & 13.485  & 6.766  & -10.750 & 13.495  & 6.741  & 0.998 & 0.999 & 1.004 \\
    \hline
    \end{tabular}
    \caption{The magnetic field components on the plane that is 12 mm away from an N52 NdFeB magnet cube described in Fig.\,\ref{fig:temp_drift} under a room temperautre of \SI{22.9}{\celsius} and \SI{25.7}{\celsius}. Column 7-9 shows the corresponding ratio between the two cases.}
    \label{tab:temp_drift}
\end{table*}

%% file: mainbody.bbl
\begin{thebibliography}{10}
\providecommand{\url}[1]{#1}
\csname url@samestyle\endcsname
\providecommand{\newblock}{\relax}
\providecommand{\bibinfo}[2]{#2}
\providecommand{\BIBentrySTDinterwordspacing}{\spaceskip=0pt\relax}
\providecommand{\BIBentryALTinterwordstretchfactor}{4}
\providecommand{\BIBentryALTinterwordspacing}{\spaceskip=\fontdimen2\font plus
\BIBentryALTinterwordstretchfactor\fontdimen3\font minus
  \fontdimen4\font\relax}
\providecommand{\BIBforeignlanguage}[2]{{%
\expandafter\ifx\csname l@#1\endcsname\relax
\typeout{** WARNING: IEEEtran.bst: No hyphenation pattern has been}%
\typeout{** loaded for the language `#1'. Using the pattern for}%
\typeout{** the default language instead.}%
\else
\language=\csname l@#1\endcsname
\fi
#2}}
\providecommand{\BIBdecl}{\relax}
\BIBdecl

\bibitem{hyperfine}
hyperfine, ``Swoop portable mr imaging system,'' \url{https://hyperfine.io/}.

\bibitem{huang2019_iMRI}
S.~Y. Huang, Z.~H. REN, S.~OBRUCHKOV, J.~GONG, R.~DYKSTRA, and W.~YU,
  ``Portable low-cost mri system based on permanent magnets/magnet arrays,''
  \emph{Investigative Magnetic Resonance Imaging {(iMRI)}}, 2019.

\bibitem{Cheng2001_CShape}
I.~Cheng, P.~J.Jungwirth, A.~J. Otter, and Y.~Wu, ``C-shaped magnetic resonance
  maging system,'' Jul. 2001, uS Patent 6,842,002 B2.

\bibitem{siemens}
{Siemens Healthcare}, ``{MAGNETOM C, 0.35 T} small footprint,''
  \url{https://www.healthcare.siemens.com/magnetic-resonance-imaging/0-35-to-1-5t-mri-scanner/magnetom-c/features}.

\bibitem{halbach1980design}
K.~Halbach, ``Design of permanent multipole magnets with oriented rare earth
  cobalt material,'' \emph{Nuclear instruments and methods}, vol. 169, no.~1,
  pp. 1--10, 1980.

\bibitem{Nishino1983singleRing}
E.~Nishino, ``Permanent magnet array for magnetic medical appliance,'' Jul.27
  1983, gB 2112645.

\bibitem{Miyajima1985Ring_pair}
G.~Miyajima, ``Cylindrical permanent magnet apparatus,'' Oct.23, formally Apr.
  4, 1984 1985, japanese Patent JPS60210804A.

\bibitem{aubert1994permanent}
G.~Aubert, ``Permanent magnet for nuclear magnetic resonance imaging
  equipment,'' Jul.~26 1994, uS Patent 5,332,971.

\bibitem{ren2018design}
Z.~H. Ren, W.~C. Mu, and S.~Y. Huang, ``Design and optimization of a ring-pair
  permanent magnet array for head imaging in a low-field portable mri system,''
  \emph{IEEE Transactions on Magnetics}, vol.~55, no.~1, pp. 1--8, 2018.

\bibitem{esparza1998low}
E.~Esparza-Coss and D.~M. Cole, ``A low cost mri permanent magnet prototype,''
  in \emph{AIP Conference Proceedings}, vol. 440, no.~1.\hskip 1em plus 0.5em
  minus 0.4em\relax AIP, 1998, pp. 119--129.

\bibitem{Danieli2009}
E.~Danieli, J.~Mauler, J.~Perlo, B.~Blümich, and F.~Casanova, ``Mobile sensor
  for high resolution nmr spectroscopy and imaging,'' \emph{J Magn Reson.},
  vol. 198, no.~1, pp. 80--, 2009.

\bibitem{OReilly2019LargeHalbach}
T.~O'Reilly, W.~Teeuwisse, L.~Winter, and A.~Webb, ``The design of a homogenous
  large-bore halbach array for low field {MRI},'' in \emph{ISMRM 2019}, 2019,
  p. 0272.

\bibitem{cooley2015}
C.~Z. Cooley, J.~P. Stockmann, B.~D. Armstrong, M.~Sarracanie, M.~H. Lev, M.~S.
  Rosen, and L.~L. Wald, ``Two-dimensional imaging in a lightweight portable
  mri scanner without gradient coils,'' \emph{Magnetic resonance in medicine},
  vol.~73, no.~2, pp. 872--883, 2015.

\bibitem{hennig2008parallel}
J.~Hennig, A.~M. Welz, G.~Schultz, J.~Korvink, Z.~Liu, O.~Speck, and
  M.~Zaitsev, ``Parallel imaging in non-bijective, curvilinear magnetic field
  gradients: a concept study,'' \emph{Magnetic Resonance Materials in Physics,
  Biology and Medicine}, vol.~21, no. 1-2, p.~5, 2008.

\bibitem{ren2015magnet}
Z.~H. Ren, L.~Mar{\'e}chal, W.~Luo, J.~Su, and S.~Y. Huang, ``Magnet array for
  a portable magnetic resonance imaging system,'' in \emph{2015 IEEE MTT-S 2015
  International Microwave Workshop Series on RF and Wireless Technologies for
  Biomedical and Healthcare Applications (IMWS-BIO)}.\hskip 1em plus 0.5em
  minus 0.4em\relax IEEE, 2015, pp. 92--95.

\bibitem{jia2019effects}
G.~Jia, S.~Y. HUANG, R.~Zhi~Hua, and W.~Yu, ``Effects of encoding fields of
  permanent magnet arrays on image quality in low-field portable mri systems,''
  \emph{IEEE Access}, vol.~9, 2019.

\bibitem{Kelton1989_multi_Rx_coil}
J.~R. Kelton, R.~L. Magin, and S.~M. Wright, ``An algorithm for rapid image
  acquisition using multiple receiver coils,'' \emph{In Proceedings of the
  ISMRM}, p. 1172, 1989.

\bibitem{cooley2015two}
C.~Z. Cooley, J.~P. Stockmann, B.~D. Armstrong, M.~Sarracanie, M.~H. Lev, M.~S.
  Rosen, and L.~L. Wald, ``Two-dimensional imaging in a lightweight portable
  mri scanner without gradient coils,'' \emph{Magnetic resonance in medicine},
  vol.~73, no.~2, pp. 872--883, 2015.

\bibitem{ren2019irregular}
Z.~H. Ren, J.~Gong, and S.~Y. Huang, ``An irregular-shaped inward-outward
  ring-pair magnet array with a monotonic field gradient for 2d head imaging in
  low-field portable {MRI},'' \emph{IEEE Access}, vol.~7, pp. 48\,715--48\,724,
  2019.

\bibitem{aubert1991cylindrical}
G.~Aubert, ``Cylindrical permanent magnet with longitudinal induced field,''
  May~7 1991, uS Patent 5,014,032.

\bibitem{zhang2017advances}
E.~Motovilova and S.~Y. Huang, ``Magnetic materials for nuclear magnetic
  resonance (nmr) and magnetic resonance imaging (mri),'' in \emph{Advances in
  Magnetic Materials: Processing, Properties and Performance}, S.~Zhang and
  D.~Zhao, Eds.\hskip 1em plus 0.5em minus 0.4em\relax CRC press, 2017, ch.~3,
  pp. 131--188.

\bibitem{cooley2021Nature_BE}
C.~Z. Cooley, P.~C. McDaniel, J.~P. Stockmann, S.~A. Srinivas, S.~F. Cauley,
  M.~Śliwiak, C.~R. Sappo, C.~F. Vaughn, B.~Guerin, M.~S. Rosen, M.~H. Lev,
  and L.~L. Wald, ``A portable scanner for magnetic resonance imaging of the
  brain,'' \emph{Nature Biomedical Engineering}, vol.~5, 2021.

\bibitem{gong2020}
J.~Gong, W.~Yu, and S.~Y. Huang, ``Image quality improvement and memory-saving
  in a permanent-magnet-array-based mri system,'' \emph{Applied Sciences},
  vol.~10, no. 2177, 2020.

\bibitem{cooley2018design}
C.~Z. Cooley, M.~W. Haskell, S.~F. Cauley, C.~Sappo, C.~D. Lapierre, C.~G. Ha,
  J.~P. Stockmann, and L.~L. Wald, ``Design of sparse halbach magnet arrays for
  portable mri using a genetic algorithm,'' \emph{IEEE transactions on
  magnetics}, vol.~54, no.~1, pp. 1--12, 2018.

\bibitem{supplement}
``See confidential supplemental material for details of derivations and
  numerical implementations.''

\bibitem{mispelter2015book}
J.~Mispelter, M.~Lupu, and A.~Briguet, \emph{NMR probeheads for biophysical and
  biomedical experiments}.\hskip 1em plus 0.5em minus 0.4em\relax Imperial
  College Press, 2015.

\bibitem{lurie2010paper}
D.~J. Lurie, S.~Aime, S.~Baroni, N.~A. Booth, L.~M. Broche, C.-H. Choi, G.~R.
  Davies, S.~Ismail, D.~{\'O}~h{\'O}g{\'a}in, and K.~J. Pine, ``Fast
  field-cycling magnetic resonance imaging,'' \emph{Comptes Rendus Physique},
  vol.~11, no.~2, pp. 136--148, 2010.

\bibitem{gong2019local_k}
J.~Gong, S.~Y. Huang, Z.~H. Ren, and W.~Yu, ``Effects of encoding fields of
  permanent magnet arrays on image quality in low-field portable {MRI}
  systems,'' \emph{IEEE Access}, 2019.

\bibitem{Furlani2001book}
E.~P. Furlani, \emph{Permanent magnet and electromechanical devices: materials,
  analysis, and applications}.\hskip 1em plus 0.5em minus 0.4em\relax Academic
  press, 2001.

\bibitem{tommy2012coil}
J.T.Vaughan, \emph{RF coils for MRI}.\hskip 1em plus 0.5em minus 0.4em\relax
  John Wiley \& Sons, 2012.

\end{thebibliography}
